\documentclass[a4paper,11pt]{article}
\pdfoutput=1 

\usepackage{jcappub} 

\usepackage[T1]{fontenc} 

\usepackage{booktabs}
\usepackage{adjustbox}
\usepackage[table,usenames,dvipsnames]{xcolor}
\usepackage{subcaption}

\newcommand{\lcdm}{$\Lambda$CDM}
\newcommand{\keq}{$k_\textrm{eq}$}
\newcommand{\threextwo}{$3\times2$\,pt}
\newcommand{\sixxtwo}{$6\times2$\,pt}
\newcommand{\Neff}{$N_\textrm{eff}$}
\newcommand{\SO}{{\em Simons Observatory}}
\newcommand{\CMBS}{{\em CMB-S4}}
\newcommand{\Planck}{{\em Planck}}
\newcommand{\Euclid}{{\em Euclid}}

\title{Improving CMB constraints on early Universe physics with LSS: A multi-probe forecast including cross-covariance}

\author[1]{Raphaël Kou\note{Corresponding author.}}
\author{and Antony Lewis}
\affiliation{Department of Physics \& Astronomy, University of Sussex, Brighton BN1 9QH, UK}

\emailAdd{r.kou@sussex.ac.uk}
\emailAdd{antony@cosmologist.info}

\abstract{Extensions to the \lcdm\ model prior to recombination can modify the growth of perturbations around radiation-matter equality, leaving a distinct signature in the matter power spectrum. Upcoming large-scale structure surveys will be sensitive to these features, allowing tests of early physics that are complementary to the CMB observations. In this paper, we forecast how well the combination of galaxy clustering, weak lensing and CMB lensing two point statistics, also known as \sixxtwo\ analysis, will tighten constraints on extensions to the \lcdm\ model in the early Universe. We find significant improvements, in particular in the case of early dark energy, where the uncertainty on its density parameter could be divided by a factor of $3$ to $4$ when combining \Euclid\ observables with \SO\ or \CMBS, compared to using CMB observations alone. Testing for different scale cuts, we find that much of the constraining power comes from the largest scales which are less prone to systematic uncertainties. 

We take into account the most significant terms in the cross-covariance between large-scale structure tracers and CMB power spectra, which arises from gravitational lensing. Assessing the impact of this additional cross-covariance on the constraints, we find small corrections for most parameters, except for $A_s$ and $\tau$ where the lensing induced covariance leads to a more significant degradation of constraints.

This forecast analysis highlights the potential of combining CMB and galaxy survey data to test the cosmological model. In particular, early Universe physics, relevant before recombination, stands out as a promising area that benefits substantially from this approach.
}

\begin{document}
\maketitle
\flushbottom

\section{Introduction}
\label{sec:intro}

Future Cosmic Microwave Background (CMB) observatories, such as \SO~\cite{SimonsObs} and \CMBS~\cite{CMB-S4:2016ple}, will have increased sensitivity at small scales and in polarization, allowing us to test physics in the primordial Universe at an unprecedented level. Such tests of fundamental physics are crucial in the context of the Hubble tension, as modifications to the standard cosmological model in the early Universe are one possible solution. Beyond the Hubble tension, however, probing early Universe physics remains of great interest, as it allows us to test our knowledge of particle physics.

Alongside these new CMB observatories, stage IV galaxy surveys, including {\em DESI}~\cite{DESI:2016fyo}, \Euclid~\cite{Euclid:2024yrr}, and {\em LSST}~\cite{LSSTDarkEnergyScience:2018jkl}, have commenced (or are nearing the start of) their observations. These large-scale structure (LSS) surveys will map the distribution of galaxies and dark matter with exceptional detail, providing complementary information to CMB observations for constraining early Universe physics. Indeed, the addition of new particles in the early Universe would alter the shape of the matter power spectrum before matter domination. For instance, introducing a new radiation-like particle would shift the scale of matter-radiation equality, \keq, affecting both the location and shape of the matter power spectrum peak. More generally, any new particle or fluid relevant before matter domination would influence the growth of perturbations in the early Universe, leaving an imprint on the matter power spectrum at large scales. Modifications to early Universe physics would also impact Baryon Acoustic Oscillations (BAO), which encode information about primordial plasma oscillations. However, models leading to significant deviations in BAO physics are already tightly constrained by CMB experiments. Consequently, most models aiming to resolve the Hubble tension are designed to avoid altering BAO features. As a result, the primary value of incorporating LSS measurements to constrain early physics lies in their sensitivity to the overall shape of the matter power spectrum at large scales, and their ability to break degeneracies between \lcdm\ and early physics parameters, though BAO features in LSS observables may still help break degeneracies to some extent. 

In this paper, we explore how the synergy between CMB and LSS observations will improve constraints on early Universe physics. In particular, we look at the combination of angular galaxy clustering and galaxy weak lensing power spectra, as well as their cross-correlation, which is often called \threextwo\ and has become a standard analysis for current and upcoming photometric galaxy surveys. We also add three other two-point statistics: the CMB lensing convergence auto-power spectrum, and its cross-correlation with the galaxy clustering and galaxy weak lensing fields,  forming a \sixxtwo\ configuration. Using different cosmological probes not only improves constraints on theoretical models, but also enables consistency checks and more robust inferences, as each probe is affected by different systematics. Many of these systematics may be uncorrelated and therefore do not contribute to cross-correlation power spectra. As a result, comparing fits from different probe combinations can help identify potential systematics, although correlated systematics may still arise and could complicate this identification.

We consider three different theoretical models: the addition of ultra-relativistic species parameterized by \Neff~\cite{Hou13,Akita20,Froustey20,Bennett21}, the Early Dark Energy (EDE) model~\cite{Doran:2006kp,Poulin:2018cxd,Poulin23}, and the recently proposed Early Dark Fluid (EDF)~\cite{Kou:2024rvn}, which provides a general parameterization of early-time modifications. Several analyses using current data~\cite{Hill:2020osr,Smith:2020rxx,Niedermann:2020qbw,Simon:2022adh,Gsponer:2023wpm} have constrained these models, but very few~\cite{Ivanov:2020ril} show the impact of combining CMB and LSS data to test them with stage IV surveys, especially with photometric datasets. Only \Neff\ has been considered more frequently in stage IV forecasts~\cite{Cerbolini:2013uya,CORE:2016npo,Sprenger:2018tdb,Brinckmann:2018owf,Euclid:2024imf}, but most of the time in the context of neutrino physics, with most of the effort concerning the mass of neutrinos. We argue that dark radiation models are of interest in their own right, and should be studied beyond the scope of neutrino physics. Since there is almost no forecast considering early physics constraints coming from stage IV galaxy surveys, we aim at filling this gap and showing their synergy with CMB observatories. We forecast constraints on all of these models with CMB observations from \SO\ and \CMBS, and look at how those constraints improve as we add \sixxtwo\ observables from \Euclid\ and the CMB lensing convergence from each corresponding CMB experiment.

When combining CMB and LSS data, most existing analyses treat them as independent probes, neglecting their covariance. Nevertheless, this is not completely correct, as gravitational lensing induces subtle, yet non-zero correlations between CMB primary observations and LSS tracers. With the very low noise levels that future observations will reach, neglecting this covariance may lead to slight biases in the cosmological inferences, as well as misestimates of the parameters uncertainties. In this paper, we show how to take into account the most significant term of this covariance and evaluate its impact on cosmological parameter constraints.

The structure of this paper is the following: section~\ref{sec:theory} introduces the theoretical models under investigation. Section~\ref{sec:observables} describes the observables used in our forecast, while section~\ref{sec:methodology} outlines our methodology. Our results are presented in section~\ref{sec:results}, and we summarize our conclusions in section~\ref{sec:conclusion}.

\section{Theoretical models}~\label{sec:theory}

\subsection{Dark radiation}
In the standard model, there are three species of neutrinos which contribute to the total relativistic
energy density given by
\begin{align}
    \rho_R=\rho_\gamma\left(1+\frac{7}{8}\left(\frac{4}{11}\right)^{4/3}N_\textrm{eff}\right)\,,
\end{align}
where $\rho_\gamma$ is the photon energy density and $N_\textrm{eff}$ is the effective number of neutrinos. In the Standard Model of particle physics, $N_\textrm{eff}=3.044$, where the deviation from $3$ accounts for the non-instantaneous decoupling of the three neutrino species~\cite{Hannestad95,Dolgov97,Gnedin98,Mangano05,DeSalas16}. Any extra relativistic species can however increase \Neff, making it a particularly interesting and convenient model to study. Several previous analyses~\cite{Cerbolini:2013uya,CORE:2016npo,Sprenger:2018tdb,Brinckmann:2018owf,Euclid:2024imf} have considered LSS forecasts for this model, but always in the context of neutrino physics, where they varied the mass of neutrinos as well. In contrast, we focus on the case where \Neff\ includes any form of extra relativistic particles relevant in the early Universe, without specifically emphasizing neutrinos. We use CAMB~\cite{Lewis:1999bs,Howlett12} to model the CMB and LSS power spectra in the \Neff\ model, choosing $N_\textrm{eff}=3.044$ as fiducial value.

\subsection{Early Dark Energy (EDE)}

Early Dark Energy (EDE) models consist in the addition of a new scalar field, initially frozen so that its density remains constant as the Universe expands, before becoming dynamical and diluting faster than matter. These models have been studied with great interest over the last years, as they are among the best candidates to solve the Hubble tension~\cite{Poulin:2018cxd,Hill:2021yec,Poulin:2021bjr,LaPosta:2021pgm} even though they were not particularly favoured by latest data~\cite{Efstathiou24,Khalife24,Hill:2020osr,McDonough24,ACT:2025tim}. CMB observations primarily constrain the angular acoustic scale $\theta_*= r_*/D_*$, where $r_*$ is the comoving sound horizon at recombination and $D_*$ is the comoving angular diameter distance to recombination. Solutions to the Hubble tension such as EDE, by injecting extra density before the recombination, tend to decrease the sound horizon, allowing $H_0$ to increase in order to keep $\theta_*$ fixed. The sound horizon is hence used as a physical reference to infer the value of $H_0$ from CMB measurements. Interestingly, several papers~\cite{Baxter:2020qlr,Farren:2021grl,Philcox:2020xbv,Philcox:2022sgj,Kable:2024mgl,Zaborowski:2024wpo} have showed that LSS probes and in particular spectroscopic galaxy power spectra enable sound horizon-free measurements of $H_0$, making instead the matter-radiation equality scale \keq\ the physical reference used to infer $H_0$. In case those measurements confirmed the Hubble tension, while assuming the \lcdm\ model in the early Universe, this could potentially rule out EDE models, as those models may not be able to consistently change both the sound horizon and \keq\ to resolve the Hubble tension. Nevertheless, previous work has not looked at how well LSS data may be able to constrain EDE models, regardless of whether they are the correct explanation, which is what we intend to do here.

In this work, we use the most usual EDE model, sometimes referred to as ‘axion-like' EDE~\cite{Poulin:2018cxd,Smith:2019ihp}. This model introduces a new scalar field $\varphi$ with potential
\begin{align}
    V(\varphi) = m^2f^2\left[1-\cos(\theta)\right]^n\,,
\end{align}
where $m$ and $f$ are the axion mass and decay constant, $n$ is an index fixed to $3$ and $\theta=\varphi/f$. As for dark radiation, we make use of CAMB to compute the CMB and LSS power spectra in the EDE model, using the best-fit values from~\cite{Smith:2019ihp} as fiducial parameters, \textit{i.e.} $\log_{10}(z_c)=3.562$, $f_\textrm{ede}(z_c)=0.122$ and $\theta_i=2.83$, where $\theta_i$ relates to the initial value of the field, while $f_\textrm{ede}(z_c)$ is the maximum fraction of the total energy density in the EDE field, reached at redshift $z_c$.

\subsection{Early Dark Fluid (EDF)}

Our Early Dark Fluid (EDF) model~\cite{Kou:2024rvn} is a phenomenological parameterization designed to provide a unified framework for testing multiple early Universe models. This model introduces an additional fluid whose density is parameterized through four amplitudes $d_1$, $d_2$, $d_3$ and $d_4$ which multiply modes with fixed scale factor evolution specifically estimated to be well constrained by \SO\ data. In addition to these four density parameters, the EDF parameterization also includes two sound speed parameters $c_1^2$ and $c_2^2$ such that the total rest-frame sound speed of the additional fluid follows

\begin{align}
    \overline{c}_s^2(a) = 
        \begin{cases}
          c_1^2 & \text{if $a \leq a_1$} \\
          c_1^2 + (c_2^2-c_1^2)\frac{\log{a}-\log{a_1}}{\log{a_2}-\log{a_1}} & \text{if $a_1 \leq a \leq a_2$}\\
          c_2^2 & \text{if $a \geq a_2$} \,,
        \end{cases}   
\end{align}
where $a$ is the scale factor, and $(a_1,a_2)=(10^{-5},10^{-3})$. It was shown that this model was able to reproduce, to some extent, the effect of dark radiation and EDE models on CMB power spectra, and could be used as a general framework to test early Universe physics. It is therefore of interest to see how well LSS can improve constraints on this model. While it was shown that specific combinations of EDF parameters can reproduce the impact of various early Universe models on the CMB power spectra, this correspondence may not always extend to LSS observables. The phenomenological nature of the EDF means that different theoretical models can map onto similar CMB signatures within this framework, while potentially yielding different predictions for LSS. Nonetheless, the EDF remains a valuable framework as a physically consistent and flexible model that would be able to capture deviations from \lcdm\ in the early Universe. Importantly, any detection within this framework would naturally motivate testing more theoretically grounded models to better understand the underlying physics.

We use the modified CAMB version\footnote{\href{https://github.com/raphkou/CAMB_EDF}{https://github.com/raphkou/CAMB\_EDF}} provided by~\cite{Kou:2024rvn} to compute CMB and LSS power spectra. For the fiducial model, we set all the density amplitudes to $0.01$ and test two different cases for the sound speed parameters. In one case, we set $c_1^2=c_2^2=0.33$, which Ref.~\cite{Kou:2024rvn} found as a possible way to solve the Hubble tension by adding a contribution that resembles a tightly-coupled dark radiation. In a second case, we use $c_1^2=c_2^2=0.9$, which looks like adding a quintessence field. Setting these parameters slightly below $1$, rather than exactly $1$, helps ensure more stable computation of the Fisher matrix in our forecast.

\section{Observables}~\label{sec:observables}
\subsection{CMB primary}
We consider CMB primary data (TT, EE, TE but no BB power spectrum) from \SO\ and \CMBS, two experiments that will observe $40\%$ of the sky with very high precision. We also include constraining power from \Planck\ coming from larger scales that \SO\ and \CMBS\ will not be able to observe, and from parts of the sky that they will not cover.

\SO\ consists of several $0.5$-meter refracting Small Aperture Telescopes (SATs) and one Large Aperture Telescope (LAT) with a $6$-meter primary mirror. In this work, we only focus on the LAT, which is the most relevant for our science target. The LAT detectors will span six frequency bands from $27$ to $280$ GHz, with temperature noise levels and beam FWHM varying between $8$ to $71\,\mu K\textrm{arcmin}$ and $0.9$ to $7.4\,\textrm{arcmin}$. We use the foreground-cleaned, beam-corrected noise curves (at baseline level) provided publicly by the \SO\ collaboration to model the temperature and polarization power spectra noise. This is slightly different from the recent forecast~\cite{SimonsObservatory:2025wwn} performed by the \SO\ collaboration, that uses (updated) goal level noise curves. We include EE and TE power spectra multipoles between $40\leq \ell \leq 5000$ and TT multipoles between $40\leq \ell \leq 3000$ as higher TT multipoles may be affected by foreground and atmospheric contamination.

\CMBS\ is another future CMB experiment that will follow \SO\ and is designed to be the definitive ground-based CMB project. It will have exquisite sensitivity, such that we use a temperature and polarization noises of $\sigma_T=1\,\mu K \textrm{arcmin}$ and $\sigma_E=\sqrt{2}\,\mu K \textrm{arcmin}$ with a beam FWHM of $\sigma_\textrm{FWHM}=1\,\textrm{arcmin}$. We then model the noise of the temperature and polarization power spectra as
\begin{align}\label{eq:Nl}
    N_\ell^{XX} = \sigma_X^2\exp{\left(\ell(\ell+1)\sigma_\textrm{FWHM}^2/(8\ln{2})\right)}\,
\end{align}
where $X\in\{T,E\}$. As for \SO, we include EE and TE multipoles in the range $40\leq\ell\leq 5000$ but use TT multipoles only between $40\leq\ell\le3000$. In order to reduce the size of the data vector (and the covariance), we bin the power spectra on multipoles such that we use $15$ logarithmically spaced multipole bins between $40\leq\ell\leq100$ and linearly spaced multipoles between $100\leq\ell\leq5000$ with width $\Delta \ell=10$. We use the same binning scheme for both \SO\ and \CMBS.

Since both \SO\ and \CMBS\ will only observe $40\%$ of the sky and will measure power spectra at multipoles above $\ell>40$, it will be possible to include additional information obtained by \Planck~\cite{Planck:2018nkj}. The \Planck\ satellite indeed covered about $70\%$ of the sky, and was able to probe low multipoles, providing crucial information, in particular in order to constrain the reionization optical depth. We model the temperature and polarization noise curves using equation~\ref{eq:Nl}, with $\sigma_T=23\,\mu K\textrm{arcmin}$, $\sigma_E=42\,\mu K\textrm{arcmin}$ and $\sigma_\textrm{FWHM}=7\,\textrm{arcmin}$. Because of optical depth uncertainty due to systematics, several forecast analyses~\cite{Bermejo-Climent:2021jxf,Euclid:2021qvm} have inflated $N_\ell^{EE}$ at multipoles below $\ell<30$. We use an alternative approach consisting in not including polarization data at those scales, but directly adding \Planck\ prior on the optical depth $\sigma(\tau)=0.0073$~\cite{Planck18} in the Fisher forecast. We hence include TT multipoles in the range $2\leq\ell\leq2500$ and TE and EE multipoles between $30\leq\ell\leq2500$. For multipoles below $\ell<40$, we consider a sky fraction of $0.7$, whereas we use $0.3$ at multipoles above $\ell\geq 40$, reflecting a conservative approach in which we do not take into account overlapping observations between \Planck\ and \SO\ or \CMBS.

In all this forecast, we do not account for delensing of the CMB power spectra, although it has been shown to improve constraints on certain early-universe parameters~\cite{Green:2016cjr,Hotinli:2021umk,Ange:2023ygk}. Specifically, delensing is particularly effective when varying \Neff\ and the primordial helium fraction simultaneously, leading to a tightening of constraints by approximately $20\%$ for a Stage IV survey~\cite{Green:2016cjr}.

\subsection{Large-scale structure tracers}
The large-scale structure observables we consider in this analysis are the CMB lensing convergence $\kappa$, the galaxy density field $g$ and the cosmic shear field $\gamma$. This work focuses on constraints that could be obtained for a \Euclid-like experiment, so we use known \Euclid\ specifications~\cite{Euclid:2019clj,Euclid:2021qvm}, especially in terms of galaxy distribution, as explained in more detail in section~\ref{sec:galaxy_clustering}. \Euclid\ will observe about $15000\,\textrm{deg}^2$, equivalent to a sky fraction $f_\textrm{sky}=0.36$. Taking the auto-power spectrum and all possible cross-correlations between all the fields mentioned previously leads to the use of six different two point statistics $C_\ell^{gg}$, $C_\ell^{\gamma\gamma}$, $C_\ell^{\kappa\kappa}$, $C_\ell^{g\gamma}$, $C_\ell^{g\kappa}$, $C_\ell^{\gamma\kappa}$, so this combination of probes is referred to as \sixxtwo, which extends the now usual \threextwo\ analysis consisting in $C_\ell^{gg}$, $C_\ell^{\gamma\gamma}$ and $C_\ell^{g\gamma}$. We follow \Euclid\ pessimistic forecast~\cite{Euclid:2019clj} by restricting our analysis to multipoles in the range $10<\ell_\textrm{GC}<750$ for galaxy clustering and $10<\ell_\textrm{WL}<1500$ for galaxy weak lensing. We also investigate an even more conservative multipole scheme by setting $10<\ell_\textrm{GC}<500$ and $10<\ell_\textrm{WL}<750$ since we expect most constraints to come from large scales through the shape of the matter power spectrum around \keq. In the following, we refer to those two configurations as ‘pessimistic' or ‘conservative', the latter being the most restrictive. This conservative scheme is introduced primarily to assess the extent to which our constraints depend on the choice of scale cuts. As for the CMB lensing convergence power spectrum, still following~\cite{Euclid:2021qvm}, we use the multipole range $10\leq\ell\leq3000$ for both \SO\ and \CMBS. This extended multipole range is justified by the fact that CMB lensing probes higher redshifts where the matter field tends to be more linear, and that it should be affected by less systematic effects (\textit{e.g.} magnification bias and galaxy bias modelling for galaxy clustering, intrinsic alignment and shear calibration bias for weak lensing). We make use of $50$ logarithmically spaced multipole bins between $\ell=10$ and $\ell=3000$, removing high multipole bins for galaxy clustering and weak lensing power spectra according to the scale cuts mentioned previously. When using cross-correlations between different probes, we always use the most restrictive range of multipoles between the two probes. For the cross-correlations between CMB lensing and \Euclid\ observations, we assume a full overlap, consistently with~\cite{Euclid:2021qvm}, so that the sky fraction we consider is $36\%$. Similarly to previous \Euclid\ forecasts~\cite{Euclid:2019clj,Euclid:2021qvm}, we neglect the effects of baryonic feedback. Although these can be relevant for stage IV CMB lensing measurements and can lead to deviations of a few percent at high multipoles~\cite{Chung:2019bsk}, they primarily affect small scales. In contrast, our constraints on early-Universe physics are mainly driven by large-scale modes as we shall see. Nevertheless, we acknowledge that baryonic feedback could still have an indirect impact through parameter degeneracies.

All the large-scale structure observables trace the dark matter distribution and can be modelled as weighted integrals of the matter power spectrum, which in the Limber approximation~\cite{Limber}, are
\begin{align}\label{eq:cl_limber}
    C_\ell^{XY} = \int \frac{dz}{c}\frac{H(z)}{\chi^2(z)}W_X(z)W_Y(z)P_m\left(k=\frac{\ell+1/2}{\chi(z)},z\right),
\end{align}
where $X$ and $Y$ can be any of the three observables considered in this work, $c$ is the speed of light, $H(z)$ is the Hubble parameter at redshift $z$, $\chi$ is the comoving distance, $P_m$ is the matter power spectrum, and $k$ is the comoving wavenumber. The weights $W_X$, also called kernels or window functions, depend on each observable and are described in more detail in the corresponding sections. Note that on the largest scales, the Limber approximation breaks down so that for multipoles below $\ell<100$, we use the more general expression
\begin{align}\label{eq:cl_non_limber}
    C_\ell^{XY} = \frac{4\pi}{c^2}\int\frac{dk}{k}\mathcal{P}_\mathcal{R}(k)\int d\chi_1d\chi_2H(z_1)H(z_2)T_X(k,\chi_1)T_Y(k,\chi_2)j_\ell(k\chi_1)j_\ell(k\chi_2)W_X(\chi_1)W_Y(\chi_2),
\end{align}
where $\mathcal{P}_\mathcal{R}$ is the primordial curvature power spectrum, $T_X$ is the transfer function of the observable $X$, and $j_\ell$ is a  spherical Bessel function. All the angular power spectra in this work are modelled using CAMB, where the non-linear matter power spectrum makes use of HMcode~\cite{Mead:2020vgs}.

\subsubsection{CMB lensing convergence}
As they travel through the Universe, photons emitted from the last-scattering surface are deflected by the large-scale structure of the Universe, an effect known as gravitational lensing. As a result, the temperature and polarization fields we observe from the CMB have been lensed, which introduces correlations between different modes. Using those correlations allows us to reconstruct the lensing potential field projected along the line-of-sight, or equivalently, the convergence field, which is related to the lensing potential $\phi$ through $\kappa=-\nabla^2\phi/2$. The convergence power spectrum $C_\ell^{\kappa\kappa}$ is computed through equations~\ref{eq:cl_limber} or~\ref{eq:cl_non_limber} using the kernel
\begin{align}
    W_\kappa(z) = \frac{3}{2}\Omega_mH_0^2\frac{1+z}{H(z)}\frac{\chi(z)}{c}\left(1-\frac{\chi(z)}{\chi(z_*)}\right),
\end{align}
where $\Omega_m$ is the matter density, $H_0$ is the Hubble parameter today, and $z_*$ is the redshift of the last-scattering surface.
We use the noise curve, known as N0 bias, provided by \SO~\cite{SimonsObs} at baseline level, while we compute it using {\em plancklens}\footnote{\href{https://github.com/carronj/plancklens}{https://github.com/carronj/plancklens}}~\cite{Planck:2018lbu} for \CMBS.

\subsubsection{Galaxy clustering}~\label{sec:galaxy_clustering}

Following \Euclid\ forecast paper~\cite{Euclid:2019clj}, we assume an underlying true galaxy distribution  following
\begin{align}
    n(z) \propto \left(\frac{z}{z_0}\right)^2\exp{\left[-\left(\frac{z}{z_0}\right)^{3/2}\right]},
\end{align}
with $z_0=0.9/\sqrt{2}$. This true galaxy distribution is then convolved with the photometric redshift error described in~\cite{Euclid:2019clj} to get the observed galaxy distribution $n_i(z)$ per redshift bin $i$, which accounts for the photometric redshift uncertainties. Still following \Euclid\ forecast, we then use $10$ equi-populated redshift bins between $0.001<z<2.5$, each with a galaxy surface density of $\bar{n}_g=3\,\textrm{arcmin}^{-2}$. We hence include the ten galaxy auto-power spectra, but also all the cross-correlations between different redshift bins. Those cross-correlations are expected to be non-zero, as some of their galaxy distributions overlap due to the photometric redshift errors and the magnification bias (see the following discussion).

The galaxy clustering kernel $W_\textrm{GC}$ we use to model the power spectrum through equations~\ref{eq:cl_limber} and~\ref{eq:cl_non_limber} follows 
\begin{align}
    W_{\textrm{GC},i}(z) &= \frac{b_i}{\bar{n}_g}n_i(z),
\end{align}
where $n_i$ and $b_i$ are the galaxy redshift distribution and the linear galaxy bias in redshift bin $i$. As in~\cite{Euclid:2019clj}, we use one constant galaxy bias parameter per redshift bin, with a fiducial value given by $b_i=\sqrt{1+\bar{z}_i}$, where $\bar{z}_i$ in the mean redshift in the bin. In practice, we model the galaxy clustering power spectrum using CAMB, which allows to take into account additional systematic effects such as the magnification bias and the redshift space distortions (RSD). In particular, the magnification bias~\cite{Kaiser:1991qi,Broadhurst:1994qu,Villumsen:1995ar,Duncan:2013haa,Thiele:2019fcu} comes from the fact that galaxies are magnified through gravitational lensing, which alters the apparent flux and shifts some galaxies above or below the detection threshold depending on the position of foreground lenses. Modelling this magnification bias requires knowledge of the logarithmic slope of the number density of lensed galaxies $s(z)$~\cite{Matsubara:2000pr,Challinor11} defined as
\begin{align}
    s(z) = \frac{\partial \log_{10}(\bar{N}(z,m<m_*))}{\partial m_*}\,,
\end{align}
where $\bar{N}(z,m<m_*)$ is the number of galaxies at redshift $z$ with magnitudes below $m_*$. We therefore add one constant parameter per redshift bin $s_i$ that we let free in this forecast. The fiducial value of $s_i$ is obtained using the fitting formula from~\cite{Euclid:2021rez} (equation C.1). 

Finally, the shot noise associated to each galaxy auto-power spectrum is identical to all redshift bins as they are equi-populated, and is given by 
\begin{align}
    N_\ell^\textrm{GC} = \frac{1}{\bar{n}_g}.
\end{align}

\subsubsection{Galaxy weak lensing}

Similarly to CMB lensing, galaxy weak lensing is due to the gravitational lensing of background galaxies by the foreground large-scale structure of the Universe, which causes a distortion in the apparent shape of the background galaxies. The primary signal we are interested in is the cosmic shear $\gamma$, but we also account for galaxy Intrinsic Alignment (IA)~\cite{Heavens:2000ad,Catelan:2000vm,Bridle:2007ft,Hirata:2007np,Joachimi11,Joachimi:2015mma,Krause:2015jqa,Lamman:2023hsj}, a known systematic effect stemming from the fact that tidal forces due to local matter structures induce correlations in the intrinsic ellipticities of background galaxies. We take into account this effect through an IA kernel, using a slightly simplified version of the model from~\cite{Krause:2015jqa}, such that

\begin{align}
    W_{\textrm{WL},i}(z) &= (1+m_i)W_{\gamma,i}(z)+W_{\textrm{IA},i}(z)\label{eq:W_wl} \\
    W_{\gamma,i}(z) &= \frac{3}{2}\Omega_mH_0^2\frac{1+z}{H(z)}\frac{\chi(z)}{c}\int_z^{z_\textrm{max}}dz'n_i(z')\left(1-\frac{\chi(z)}{\chi(z')}\right)\\
    W_{\textrm{IA},i}(z) &= -n_i(z)A_i\frac{C_1\rho_{c,0}}{D(z)}(1+z)^{\eta_i}\,,
\end{align}
where $D(z)$ is the growth factor normalized today, $A_i$ and $\eta_i$ are two free parameters set to $1$ and $0$, respectively, in the fiducial model, $\rho_{c,0}$ is the critical density at $z=0$ and $C_1$ a constant that we fix such that $C_1\rho_{c,0}=0.0134$, following~\cite{Hirata:2003ka,Bridle:2007ft}. Equation~\ref{eq:W_wl} also includes shear calibration uncertainty through the multiplicative shear bias $m_i$, which is another free parameter in each redshift bin over which we marginalize in this forecast. Finally, we include the galaxy shape noise of each auto-power spectrum such that
\begin{align}
    N_\ell^\textrm{WL} = \frac{\epsilon^2}{\bar{n}_g},
\end{align}
with $\epsilon=0.26$, following~\cite{Euclid:2019wjj}.

\section{Covariance}
\subsection{Gaussian covariance}

We model the covariance of our power spectra using a Gaussian covariance matrix that follows
\begin{align}
    \textrm{Cov}^G(C_\ell^{AB},C_{\ell'}^{CD}) = \frac{\delta_{\ell\ell'}}{(2\ell+1)f_\textrm{sky}}\left[\left(C_\ell^{AC}+N_\ell^{AC}\right)\left(C_\ell^{BD}+N_\ell^{BD}\right)+\left(C_\ell^{AD}+N_\ell^{AD}\right)\left(C_\ell^{BC}+N_\ell^{BC}\right)\right],
\end{align}
where $(A,B,C,D) \in \{\textrm{T},\textrm{E},\kappa,\textrm{GC},\textrm{WL}\}$ and $f_\textrm{sky}$ is the observed sky fraction, being either $0.4$ for the CMB, or $0.36$ for the LSS observables. We use $0.36$ for the CMB lensing convergence in order to use the same sky fraction for all LSS probes, even though $0.4$ could be used for its auto-power spectrum. Note that a factor of $\Delta \ell$ (number of multipoles in a bin) should be included in the denominator when binning on multipoles. In this Gaussian covariance matrix, we neglect correlations between CMB primary power spectra and LSS power spectra, but we take into account the covariance that arises between those two datasets from gravitational lensing effect in section~\ref{sec:lensing_covariance}. As a result, our Gaussian covariance matrix results in two different blocks that correspond to either the CMB primary, or the \sixxtwo\ power spectra.

\subsection{Lensing induced (cross-)covariance}\label{sec:lensing_covariance}

\begin{figure}[htbp]
\centering
\includegraphics[width=0.85\textwidth]{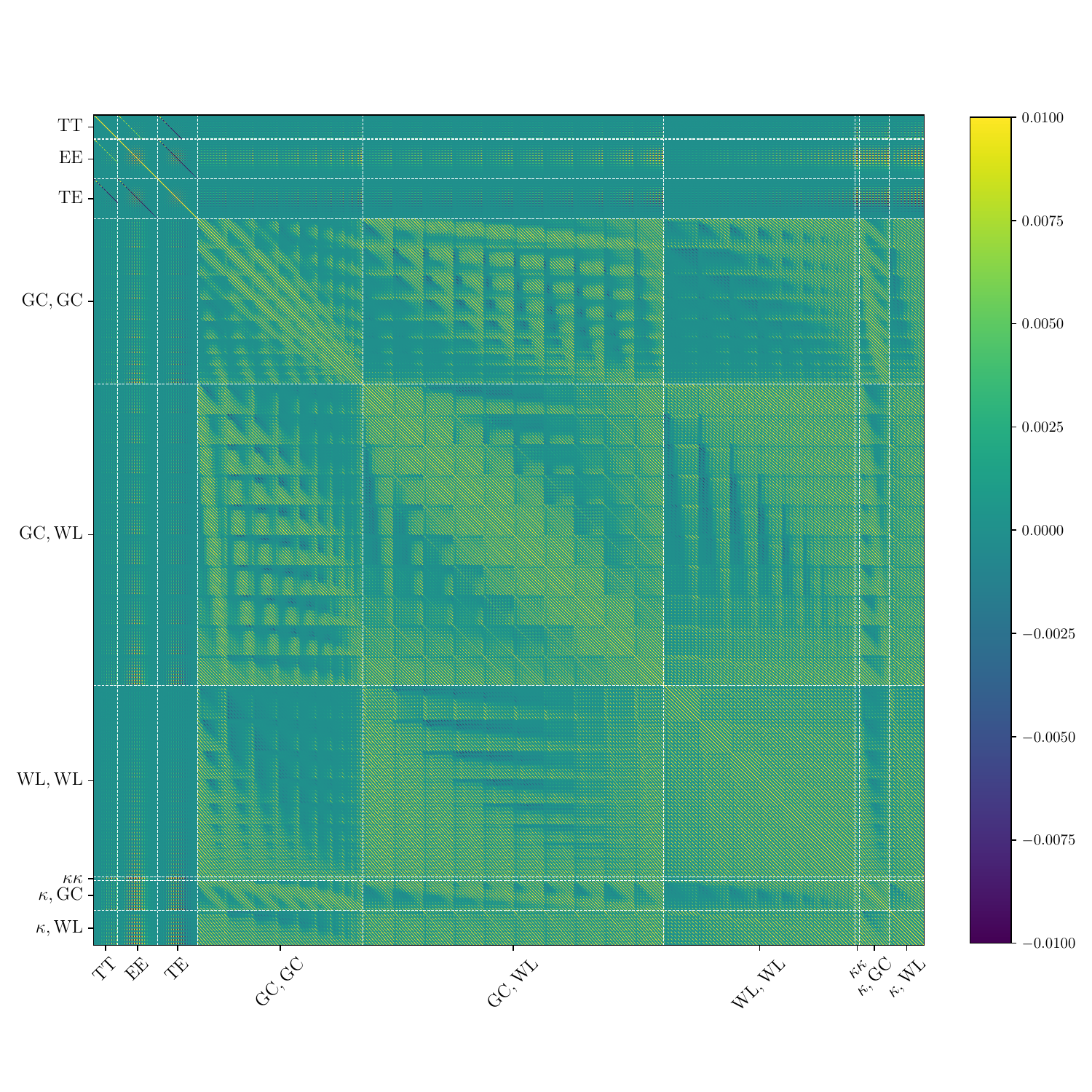}
\caption{Total correlation matrix truncated between $-0.01$ and $0.01$ including all observables considered in this forecast in the case of \CMBS\ and \Euclid. For each type of observable, the elements of our data vector are ordered first by redshift bin (if relevant) and then by multipole. Specifically, for each redshift bin $z_i$, we include the power spectrum $C_\ell(z_i)$ for all considered multipoles $\ell$, before moving to the next redshift bin. The white dashed lines are only shown to help the reader distinguishing the different blocks in the correlation matrix. \label{fig:cov_CMB_6x2}}
\end{figure}

\begin{figure}[htbp]
\centering
\includegraphics[width=\textwidth]{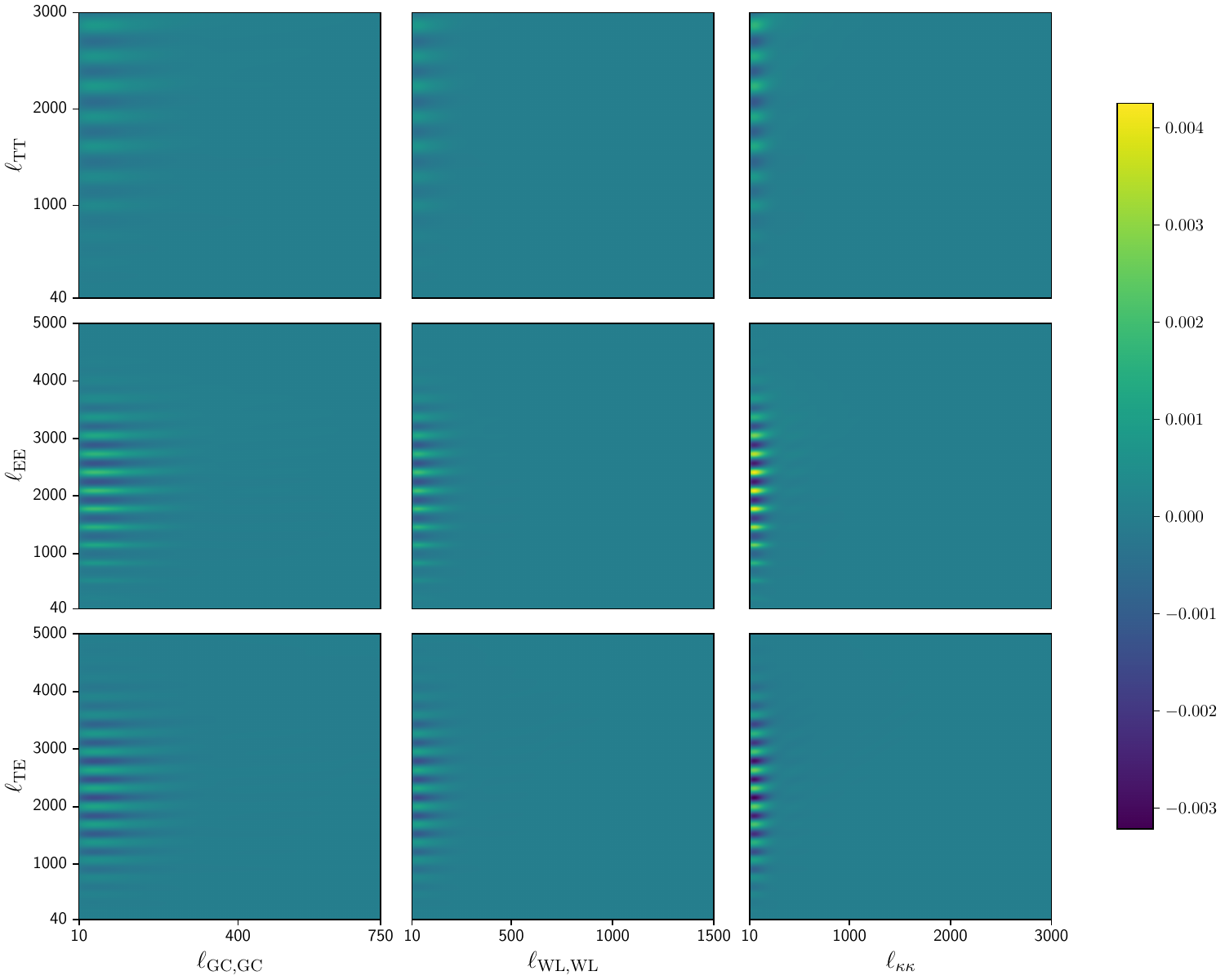}
\caption{Unbinned correlation matrix between CMB primary temperature and polarization power spectra of \CMBS, and large-scale structure probes, namely galaxy clustering, galaxy weak lensing and CMB lensing convergence power spectra. Galaxy clustering and galaxy weak lensing redshift bins used for this plot are the highest \Euclid\ redshift bin (with mean redshift $\bar{z}=1.8$), where the correlations with CMB power spectra are the most significant. The correlations we observe result from gravitational lensing, which smooths the acoustic peaks of the CMB at small scales. As a result, the correlated pattern reflects the oscillatory structure of the acoustic peaks, modulated by the large-scale lensing potential. This process induces correlations between low multipoles in the LSS probes and high multipoles in the primary CMB. However, at even higher CMB multipoles, the correlations weaken, primarily because of the increasing noise in the Gaussian covariance.}\label{fig:cross_covariance}
\end{figure}

\begin{figure}[htbp]
\centering
\includegraphics[width=0.8\textwidth]{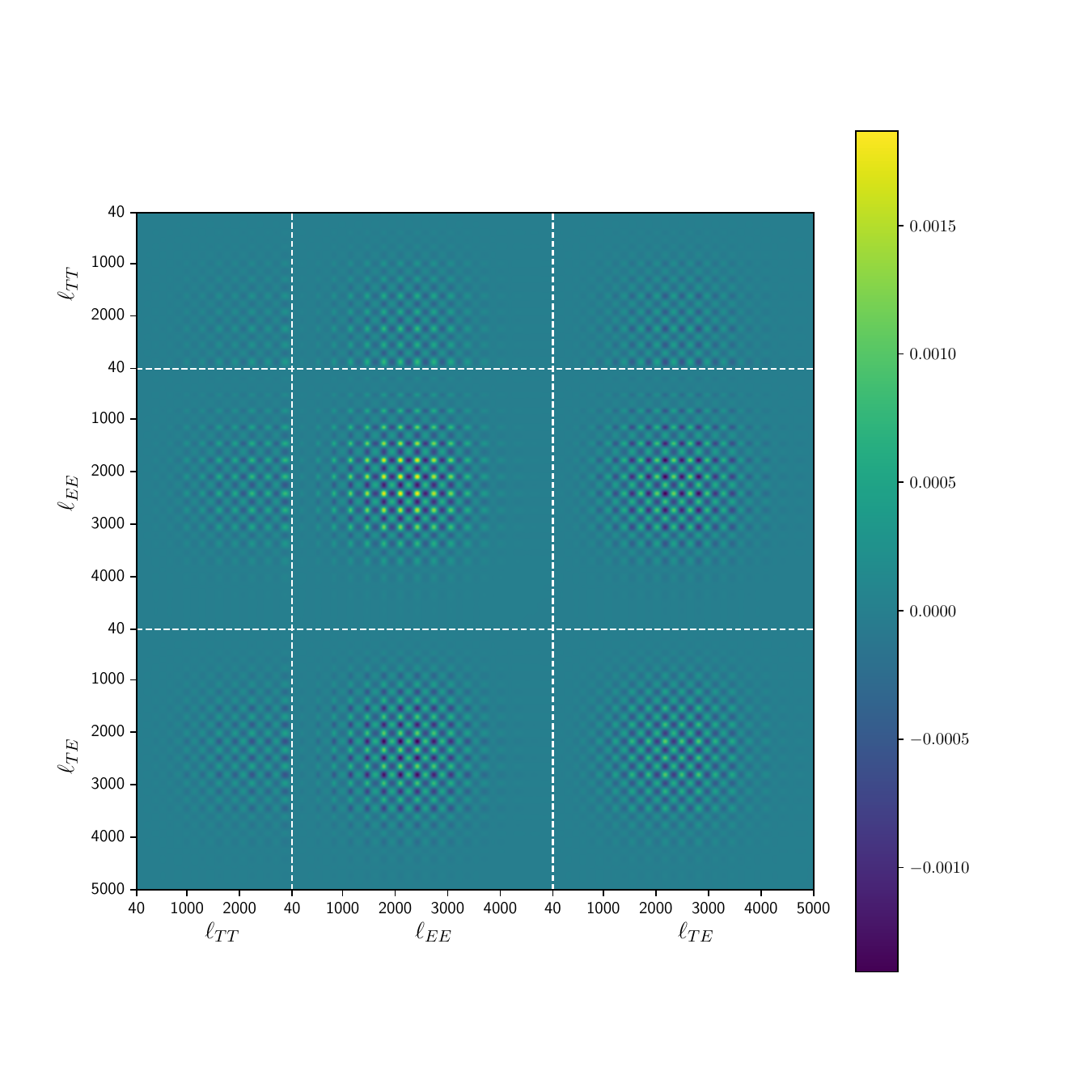}
\caption{CMB covariance matrix induced by gravitational lensing, in the case of \CMBS, where each element is normalized by the product of the square root of the corresponding diagonal elements in the total covariance matrix. This can be seen as the CMB correlation matrix where the Gaussian elements have been removed. This matrix is not truncated so that the level of correlations induced by gravitational lensing can be read from this figure. As discussed previously, the observed correlations result from lensing mixing information from different scales, thereby correlating distinct multipoles.}\label{fig:cov_CMB_lensing}
\end{figure}

As mentioned in the previous section, we estimate the covariance that arises from gravitational lensing. This introduces additional covariance in the CMB temperature and polarization block, as different modes become correlated, but it also introduces correlations between CMB and \sixxtwo\ power spectra, since a change in the lensing potential would affect both the foreground lenses and the background CMB observations. We only model the most significant term of this covariance, following~\cite{Peloton:2016kbw} to which we refer for a more comprehensive study.

The dominant term in the CMB temperature and polarization block is given by
\begin{align}
    \textrm{Cov}^L(C_{\ell_1}^{AB},C_{\ell_2}^{CD})&= \sum_{\ell_3}\frac{\partial C_{\ell_1}^{AB}}{\partial C_{\ell_3}^{\phi\phi}}\textrm{Cov}(C_{\ell_3}^{\phi\phi},C_{\ell_3}^{\phi\phi})\frac{\partial C_{\ell_2}^{CD}}{\partial C_{\ell_3}^{\phi\phi}}\\
    &=\sum_{\ell_3}\frac{\partial C_{\ell_1}^{AB}}{\partial C_{\ell_3}^{\phi\phi}}\frac{2}{(2\ell_3+1)f_\textrm{sky}}(C_{\ell_3}^{\phi\phi})^2\frac{\partial C_{\ell_2}^{CD}}{\partial C_{\ell_3}^{\phi\phi}},\label{eq:cov_lensing_cmb}
\end{align}
where $A$, $B$, $C$ and $D$ are either T or E. The derivatives of the CMB power spectra with respect to the lensing potential are estimated using CAMB. Similarly, the most significant term in the cross-covariance between CMB and LSS power spectra follows
\begin{align}
    \textrm{Cov}^L(C_{\ell_1}^{AB},C_{\ell_2}^{CD}) &= \sum_{\ell_3} \frac{\partial{C_{\ell_1}^{AB}}}{\partial C_{\ell_3}^{\phi\phi}}\textrm{Cov}(C_{\ell_3}^{\phi\phi},C_{\ell_2}^{CD})\label{eq:xcov_1} \\
    &= \sum_{\ell_3} \frac{\partial{C_{\ell_1}^{AB}}}{\partial C_{\ell_3}^{\phi\phi}}\frac{2\delta_{\ell_2\ell_3}}{(2\ell_2+1)f_\textrm{sky}}C_{\ell_2}^{\phi C}C_{\ell_2}^{\phi D}\label{eq:xcov_2} \\
    &= \frac{\partial{C_{\ell_1}^{AB}}}{\partial C_{\ell_2}^{\phi\phi}}\frac{2}{(2\ell_2+1)f_\textrm{sky}}C_{\ell_2}^{\phi C}C_{\ell_2}^{\phi D},\label{eq:xcov_3}
\end{align}
where $(A,B)\in\{\textrm{T},\textrm{E}\}$, while $(C,D)\in\{\kappa,\textrm{GC}, \textrm{WL}\}$. Equation~\ref{eq:xcov_1} contains the most significant contribution to the covariance between observables $C_{\ell_1}^{AB}$ and $C_{\ell_2}^{CD}$, where $\textrm{Cov}(C_{\ell_3}^{\phi\phi},C_{\ell_2}^{CD})$ encapsulates all correlations between $C_{\ell_2}^{CD}$ and $C_{\ell_3}^{\phi\phi}$. In equation~\ref{eq:xcov_2}, however, we approximate this covariance by assuming that $C_{\ell_3}^{\phi\phi}$ and $C_{\ell_2}^{CD}$ are only correlated through their Gaussian covariance, thereby neglecting the lensing-induced corrections to $C_{\ell_2}^{CD}$ (e.g., magnification bias, post-Born corrections, etc.).
In particular, when $C=D=\kappa$, this approximation neglects contributions from the $N_1$ bias, which arises from higher-order correlations between the reconstructed lensing field and large-scale structure. Since $N_1$ is correlated with the true lensing potential $\phi$, our approximation does not account for its contribution to the covariance between the reconstructed $C_\ell^{\kappa\kappa}$ and $C_\ell^{\phi\phi}$. We refer the reader to~\cite{Peloton:2016kbw,Lemos:2023xhs} for a more careful treatment of the cross-covariance between CMB primary and lensing power spectra, taking into account the $N_1$ reconstruction bias. The value of $f_\textrm{sky}$ we use in this cross-covariance is the one of \Euclid\ as it is the smallest of the two datasets. This is consistent with the assumption of full overlap between CMB and LSS observations. When considering multipole bins, both equations~\ref{eq:cov_lensing_cmb} and~\ref{eq:xcov_3} must be averaged over the corresponding band powers. Alternatively, averaging equation~\ref{eq:xcov_3} over CMB multipoles while using the central multipole of the LSS bin provides a very good approximation. For instance, in the case of the cross-covariance between CMB lensing and the CMB primary power spectra, the relative error introduced by this approximation is at the percent level for most of the relevant scales ($\ell_{\kappa\kappa} < 200$, see figure~\ref{fig:cross_covariance}) given the binning scheme we considered.

Figure~\ref{fig:cov_CMB_6x2} shows the total correlation matrix (truncated between $-0.01$ and $0.01$) we computed in this analysis, including the CMB temperature and polarization power spectra in the \CMBS\ case, as well as the \sixxtwo\ observables for the combination of \Euclid\ and \CMBS. The correlation matrix is defined such that
\begin{align}\label{eq:corr}
    \textrm{Corr}_{ij}=\frac{\textrm{Cov}_{ij}}{\sqrt{\textrm{Cov}_{ii}\textrm{Cov}_{jj}}}.
\end{align}
Similarly, figure~\ref{fig:cross_covariance} and~\ref{fig:cov_CMB_lensing}, respectively, show the CMB-LSS cross-covariance and the correlation matrix of the lensing induced covariance matrix (normalized by the total covariance matrix, \textit{i.e.} the numerator in equation~\ref{eq:corr} is the lensing induced covariance matrix while the denominator uses the total covariance matrix) for the CMB-only block, in the case of \CMBS. Those two figures are shown in the multipole-unbinned case so that the correlation levels are meaningful and do not depend on the binning scheme. As shown in~\cite{Peloton:2016kbw}, the cross-covariance between CMB and LSS probes mostly relates large LSS scales to small CMB scales. This is expected, as the main effect of gravitational lensing onto CMB auto-power spectra is a smoothing of the small scale acoustic peaks due to the mix of information that large-scale structures induce. At even smaller (CMB) scales, the correlation coefficients corresponding to the cross-covariance decrease, which is due to the rapidly increasing noise that contributes to the Gaussian covariance. Likewise, the contribution of the lensing covariance to the total covariance in the CMB-only block (figure~\ref{fig:cov_CMB_lensing}) is always a rather small correction of the order of $10^{-3}$ that mainly arises at small scales, while even smaller scales ($\ell \sim 4000-5000$) are mainly affected by the noise increase. We show these correlation matrices in the case of \SO\ in appendix~\ref{appendix:SO_cov} (figures~\ref{fig:cross_covariance_SO} and~\ref{fig:cov_CMB_lensing_SO}). Compared to the case of \CMBS, the matrices exhibit the same structure but show weaker correlations, with overall strengths reduced by approximately a factor of 2. The highest correlations also shift toward slightly larger scales, as noise in \SO\ starts to dominate at larger scales, increasing its contribution to the Gaussian covariance compared to \CMBS.

\section{Methodology}~\label{sec:methodology}~We use all the recipes described previously to perform a Fisher forecast whose formalism is summarized here. The Fisher information matrix~\cite{Carron13,Bellomo:2020pnw,Euclid:2021qvm} is given by
\begin{align}\label{eq:fisher}
    F_{\alpha,\beta} = \left(\frac{\partial \mathcal{D}}{\partial \alpha}\right)^T\left(\textrm{Cov}^G+\textrm{Cov}^L\right)^{-1}\frac{\partial \mathcal{D}}{\partial \beta},
\end{align}
where $\alpha$ and $\beta$ are two of the parameters varied in the forecast, and $\mathcal{D}$ denotes our full data vector consisting in $C_\ell^{TT}$, $C_\ell^{EE}$ and $C_\ell^{TE}$, as well as in the \sixxtwo\ power spectra, including all possible cross-correlations between redshift bins. We similarly compute a second Fisher matrix for \Planck, which we then add to the first Fisher matrix. This means we neglect all correlations between \Planck\ and the other experiments, which we justify by the fact that we only included \Planck\ data at multipoles or sky areas that will not overlap with the other experiments. There is however some overlap between \Planck\ low multipoles and the LSS tracers, but the cross-covariance due to lensing is extremely small at those scales, and we neglect the covariance due to the Integrated Sachs Wolfe (ISW) effect, as discussed in more detail at the end of this section. The 1-$\sigma$ uncertainty in a parameter $\alpha$ can then be estimated after marginalizing over all the other parameters in the analysis through
\begin{align}\label{eq:sigma_x}
    \sigma(\alpha) = \sqrt{\left(F^{-1}\right)_{\alpha,\alpha}}.
\end{align}
We summarize here the key points of our analysis. We test three different early Universe physics models, namely a dark radiation model parameterized through \Neff, the axion-like EDE and the EDF model (with a dark radiation-like and a quintessence-like setups). For each of those models, we use the Fisher formalism to predict the constraints that will be obtained by \SO\ and \CMBS, using multipoles in the range $40\leq\ell\leq 5000$ for $C_\ell^{EE}$ and $C_\ell^{TE}$, $40\leq\ell\leq 3000$ for $C_\ell^{TT}$ with a sky coverage of $40\%$. We combine those future CMB observations with \Planck\ data, including all the $C_\ell^{TT}$ multipoles, as well as $C_\ell^{TE}$ and $C_\ell^{EE}$ for multipoles between $30\leq\ell\leq2500$. Low polarization multipoles are replaced by a Gaussian prior on the optical depth such that $\sigma(\tau)=0.0073$. At low multipoles, we consider a sky coverage of $70\%$, but only $30\%$ when \Planck\ multipoles overlap with the ones from \SO\ or \CMBS.
We then add LSS data to assess how those observations will improve the constraints on early Universe physics. Focusing on an \Euclid-like experiment, we consider $10$ redshift bins in which galaxy clustering and galaxy weak lensing power spectra will be measured, each with a $3\,\textrm{arcmin}^{-2}$ galaxy surface density. We also add CMB lensing power spectrum measurements from \SO\ or \CMBS, and all the cross-correlations between those three observables, yielding the so-called \sixxtwo\ analysis. We model the galaxy clustering power spectra using one linear galaxy bias $b_i$ per redshift bin, and the magnification bias with one $s_i$ parameter per bin as well. We include multiplicative shear bias and intrinsic alignment systematic effects, requiring to add extra $m_i$, $A_i$ and $\eta_i$ parameters in each redshift bin. As a result, we have between $57$ and $62$ free parameters, depending on the theoretical model we test. The multipole range we use for the CMB lensing power spectrum is always $10\leq\ell\leq3000$, whereas we use $10\leq\ell_\textrm{GC}\leq750$ and $10\leq\ell_\textrm{WL}\leq1500$, or $10\leq\ell_\textrm{GC}\leq500$ and $10\leq\ell_\textrm{WL}\leq750$ as a more conservative multipole range for the galaxy and weak lensing power spectra. We chose to test the case of very restricted multipole ranges for galaxy and weak lensing power spectra because they may be affected by more systematic effects (magnification bias, calibration, intrinsic alignment) than the CMB lensing power spectrum, and because the galaxy bias linearity assumption may break down at small scales. We also test different combinations of observables, not always including the six LSS power spectra, in order to quantify the improvement due to specific combinations of probes.

We model the covariance as the sum of a Gaussian covariance matrix, and a lensing-induced covariance matrix which impacts both the CMB-only block, and the cross-covariance between CMB and LSS power spectra. We also quantify the constraints that would be obtained using the Gaussian covariance only, so that we can assess the impact of including the lensing covariance in the cosmological constraints.

Even if we took into account most of the known systematics, there remains a few effects that were not included in this analysis and that could be improved. In particular, it has been shown that the inclusion of the Super Sample Covariance (SSC)~\cite{Hu:2002we,Lacasa:2016yva,Beauchamps:2021fhb,Euclid:2023ove}, which arises from the coupling of supersurvey modes with small scale modes, led to a severe degradation of the \Euclid\ constraining power. As will be shown in the following section, the improvement we obtain when adding LSS tracers to CMB observations mainly comes from the addition of large scales, especially in the case of EDE, where SSC is subdominant. It is therefore likely that SSC may not affect much those constraints, even if it could still have an impact in the case of dark radiation. Similarly, we neglected the cross-covariance between CMB temperature and LSS tracers coming from the ISW effect, especially when adding \Planck\ low multipoles. We believe that this covariance should have limited impact on this forecast, as we anyway found that adding \Planck\ low multipoles did not improve the constraints much.

\begin{table}[ht]
    \centering
    \resizebox{0.95\textwidth}{!}{
    \begin{tabular}{lcccccccccccc}
    \hline
        \rowcolor{Emerald!30}
        \multicolumn{13}{c}{\textbf{Dark radiation (\Neff)}} \\
        \hline
        & $\theta_s$ & $\omega_b$ &$\omega_c$ & $\tau$ & $A_s$ & $n_s$ & \Neff & & & & & \\
        \hline
        \rowcolor{gray!15} 
        \textbf{\SO} & $\times10^{-6}$& $\times10^{-5}$ & $\times10^{-3}$ & $\times10^{-3}$ & $\times10^{-11}$ & $\times10^{-3}$ & $\times10^{-2}$ & & & & & \\
        CMB only & $1.74$ & $8.17$ & $1.25$ & $5.80$ & $2.45$ & $3.97$ & $6.46$ & & & & & \\
        + $C_\ell^{\kappa\kappa}$ & $1.58$ & $8.10$ & $0.990$ & $5.43$&$2.11$ &$3.84$ & $5.86$ & & & & & \\
        + \threextwo\ pess. & $0.911$ & $6.88$ & $0.618$& $4.43$ & $1.76$ & $2.52$ & $3.81$ & & & & & \\
        + \threextwo\ cons. & $1.04$ & $7.40$ & $0.720$ & $4.98$ & $2.05$& $3.11$&$4.74$ & & & & & \\
        + \sixxtwo\ pess. & $0.875$&$6.84$ &$0.610$ & $3.90$& $1.49$&$2.38$ &$3.77$ & & & & & \\
        + \sixxtwo\ cons. & $1.01$ & $7.33$& $0.699$& $4.23$&$1.68$ & $3.02$& $4.66$& & & & & \\
        \hline
        \rowcolor{gray!15} 
        \textbf{\CMBS} & $\times10^{-6}$& $\times10^{-5}$ & $\times10^{-4}$ & $\times10^{-3}$ & $\times10^{-11}$ & $\times10^{-3}$ & $\times10^{-2}$ & & & & & \\
        CMB only & $1.12$ & $3.86$ & $8.95$ & $5.04$ & $1.97$ & $3.02$ & $3.97$ & & & & & \\
        + $C_\ell^{\kappa\kappa}$ & $0.971$ & $3.74$ & $6.27$ & $4.88$ & $1.79$ & $2.78$ & $3.10$ & & & & & \\
        + \threextwo\ pess. & $0.692$ & $3.57$ & $4.84$ & $3.84$ & $1.51$ & $2.18$ & $2.59$ & & & & & \\
        + \threextwo\ cons. & $0.753$ & $3.70$ & $5.49$ & $4.22$ & $1.69$ & $2.57$ & $3.11$ & & & & & \\
        + \sixxtwo\ pess. & $0.645$ & $3.54$ & $4.40$ & $3.03$ & $1.09$ & $2.05$ & $2.50$ & & & & & \\
        + \sixxtwo\ cons. & $0.684$ & $3.63$ & $4.74$ & $3.27$ & $1.19$ & $2.40$ & $2.82$ & & & & & \\
        \hline
        \rowcolor{Emerald!30}
        \multicolumn{13}{c}{\textbf{Early Dark Energy}} \\
        \hline
        & $\theta_s$ & $\omega_b$ &$\omega_c$ & $\tau$ & $A_s$ & $n_s$ & $\theta_i$ & $\log_{10}{(z_c)}$ & $f_\textrm{ede}(z_c)$ & & & \\
        \hline
        \rowcolor{gray!15} 
        \textbf{\SO} & $\times10^{-6}$& $\times10^{-5}$ & $\times10^{-3}$ & $\times10^{-3}$ & $\times10^{-11}$ & $\times10^{-3}$ & $\times10^{-2}$ & $\times10^{-2}$ & $\times10^{-2}$ & & & \\
        CMB only & $1.61$& $8.06$& $2.65$& $5.77$&$2.52$ &$5.10$ & $4.12$& $2.02$&$2.23$ & & & \\
        + $C_\ell^{\kappa\kappa}$ & $1.47$ & $7.83$ & $2.07$ & $5.58$ & $2.28$ & $4.57$ & $4.03$ & $1.96$ & $1.88$ & & & \\
        + \threextwo\ pess. & $1.23$ & $6.13$ & $0.871$ & $4.52$ & $1.83$ & $2.39$ & $3.85$ & $1.73$ & $0.510$ & & & \\
        + \threextwo\ cons. & $1.25$ & $6.14$ & $0.952$ & $4.89$& $2.01$ & $2.47$ & $3.87$ & $1.75$ & $0.555$ & & & \\
        + \sixxtwo\ pess. & $1.22$ & $6.07$ & $0.842$ & $3.90$ & $1.47$ & $2.26$ & $3.84$ & $1.69$ & $0.488$ & & & \\
        + \sixxtwo\ cons. & $1.24$ & $6.08$ & $0.909$ & $4.11$ & $1.56$ & $2.37$ & $3.86$ & $1.71$ & $0.524$ & & & \\
        \hline
        \rowcolor{gray!15} 
        \textbf{\CMBS} & $\times10^{-6}$& $\times10^{-5}$ & $\times10^{-3}$ & $\times10^{-3}$ & $\times10^{-11}$ & $\times10^{-3}$ & $\times10^{-2}$ & $\times10^{-2}$ & $\times10^{-2}$ & & & \\
        CMB only & $1.14$ & $4.00$ & $1.74$ & $5.28$ & $2.20$ & $3.67$ & $3.82$ & $1.68$ & $1.46$ & & & \\
        + $C_\ell^{\kappa\kappa}$ & $1.05$ & $3.71$ & $1.18$ & $5.20$ & $2.10$ & $3.02$ & $3.71$ & $1.64$ & $1.09$ & & & \\
        + \threextwo\ pess. & $0.974$ & $3.20$ & $0.820$ & $3.72$ & $1.48$ & $2.22$ & $3.57$ & $1.41$ & $0.481$ & & & \\
        + \threextwo\ cons. & $0.985$ & $3.22$ & $0.893$ & $3.91$ & $1.56$ & $2.31$ & $3.59$ & $1.44$ & $0.522$ & & & \\
        + \sixxtwo\ pess. & $0.951$ & $3.16$ & $0.751$ & $2.96$ & $1.03$ & $2.06$ & $3.51$ & $1.36$ & $0.437$ & & & \\
        + \sixxtwo\ cons. & $0.965$ & $3.18$ & $0.795$ & $3.14$ & $1.09$ & $2.16$ & $3.53$ & $1.39$ & $0.459$ & & & \\
        \hline
        \rowcolor{Emerald!30}
        \multicolumn{13}{c}{\textbf{Early Dark Fluid ($c_1^2=c_2^2=0.33$, $d_1=d_2=d_3=d_4=0.01$)}} \\
        \hline
        & $\theta_s$ & $\omega_b$ &$\omega_c$ & $\tau$ & $A_s$ & $n_s$ & $d_1$ & $d_2$ & $d_3$ & $d_4$ & $c_1^2$ & $c_2^2$ \\
        \hline
        \rowcolor{gray!15} 
        \textbf{\SO} & $\times10^{-6}$& $\times10^{-5}$ & $\times10^{-3}$ & $\times10^{-3}$ & $\times10^{-11}$ & $\times10^{-3}$ & $\times10^{-3}$ & $\times10^{-3}$ & $\times10^{-3}$ & $\times10^{-3}$ & $\times10^{-1}$ & $\times10^{-2}$ \\
        CMB only & $2.97$ & $7.86$ & $1.67$ & $6.45$ & $2.53$ & $5.65$ & $5.54$ & $4.13$ & $4.45$ & $6.04$ & $1.47$ & $7.10$ \\
        + $C_\ell^{\kappa\kappa}$ & $2.89$ & $7.83$ & $1.62$ & $6.31$ & $2.41$ & $5.18$ & $5.53$ & $3.98$ & $4.41$ & $5.68$ & $1.31$ & $6.24$ \\
        + \threextwo\ pess. & $2.43$ & $7.26$ & $0.785$ & $5.25$ & $1.95$ & $3.22$ & $3.27$ & $2.90$ & $3.19$ & $4.27$ & $0.739$ & $4.55$ \\
        + \threextwo\ cons. & $2.57$ & $7.30$ & $0.861$ & $5.59$ & $2.13$ & $3.87$ & $3.43$ & $2.99$ & $3.47$ & $4.77$ & $0.910$ & $4.95$ \\
        + \sixxtwo\ pess. & $2.40$ & $7.21$ & $0.694$ & $4.67$ & $1.68$ & $3.05$ & $3.13$ & $2.85$ & $3.14$ & $4.16$ & $0.709$ & $4.50$ \\
        + \sixxtwo\ cons. & $2.54$ & $7.25$ & $0.778$ & $4.88$ & $1.79$ & $3.73$ & $3.29$ & $2.93$ & $3.41$ & $4.68$ & $0.886$ & $4.92$ \\
        \hline
        \rowcolor{gray!15} 
        \textbf{\CMBS} & $\times10^{-6}$& $\times10^{-5}$ & $\times10^{-3}$ & $\times10^{-3}$ & $\times10^{-11}$ & $\times10^{-3}$ & $\times10^{-3}$ & $\times10^{-3}$ & $\times10^{-3}$ & $\times10^{-3}$ & $\times10^{-2}$ & $\times10^{-2}$ \\
        CMB only & $1.84$ & $4.93$ & $1.25$ & $6.13$ & $2.36$ & $4.39$ & $3.74$ & $3.32$ & $4.16$ & $4.49$ & $9.22$ & $4.14$ \\
        + $C_\ell^{\kappa\kappa}$ & $1.74$ & $4.88$ & $1.10$ & $5.96$ &  $2.29$ & $3.83$ & $3.50$ & $3.22$ & $4.00$ & $4.11$ & $8.14$ & $3.69$ \\
        + \threextwo\ pess. & $1.54$ & $4.56$ & $0.715$ & $4.80$ & $1.74$ & $2.94$ & $2.76$ & $2.58$ & $2.88$ & $3.56$ & $6.35$ & $3.13$ \\
        + \threextwo\ cons. & $1.65$ & $4.60$ & $0.764$ & $5.01$ & $1.85$ & $3.49$ & $2.85$ & $2.66$ & $3.16$ & $3.94$ & $7.52$ & $3.46$ \\
        + \sixxtwo\ pess. & $1.49$ & $4.49$ & $0.602$ & $3.80$ & $1.30$ & $2.71$ & $2.57$ & $2.44$ & $2.74$ & $3.36$ & $5.89$ & $3.02$ \\
        + \sixxtwo\ cons. & $1.60$ & $4.54$ & $0.658$ & $3.99$ & $1.38$ & $3.20$ & $2.66$ & $2.51$ & $2.98$ & $3.71$ & $6.98$ & $3.31$ \\
    \hline
    \rowcolor{Emerald!30}
        \multicolumn{13}{c}{\textbf{Early Dark Fluid ($c_1^2=c_2^2=0.9$, $d_1=d_2=d_3=d_4=0.01$)}} \\
        \hline
        & $\theta_s$ & $\omega_b$ &$\omega_c$ & $\tau$ & $A_s$ & $n_s$ & $d_1$ & $d_2$ & $d_3$ & $d_4$ & $c_1^2$ & $c_2^2$ \\
        \hline
        \rowcolor{gray!15} 
        \textbf{\SO} & $\times10^{-6}$& $\times10^{-5}$ & $\times10^{-3}$ & $\times10^{-3}$ & $\times10^{-11}$ & $\times10^{-3}$ & $\times10^{-3}$ & $\times10^{-3}$ & $\times10^{-2}$ & $\times10^{-2}$ &  & $\times10^{-1}$ \\
        CMB only & $5.03$ & $9.31$  & $2.94$  & $6.52$ &  $2.50$ & $5.64$ & $8.27$ & $5.40$ & $1.43$ & $1.27$ & $1.03$ & $6.36$ \\
        + $C_\ell^{\kappa\kappa}$ & $4.57$ & $8.75$ & $2.36$ & $6.41$ & $2.42$ & $5.30$ & $8.05$ & $5.32$ & $1.16$ & $1.23$ & $0.920$ & $6.17$ \\
        + \threextwo\ pess. & $3.52$ & $6.99$ & $0.742$ & $5.32$ & $1.96$ & $3.37$ & $6.66$ & $4.19$ & $0.545$ & $0.980$ & $0.738$ & $3.72$ \\
        + \threextwo\ cons. & $3.83$ & $7.05$ & $0.776$ & $5.67$ & $2.13$ & $3.69$ & $7.00$ & $4.24$ & $0.559$ & $1.07$ & $0.783$ & $3.87$ \\
        + \sixxtwo\ pess. & $3.47$ & $6.95$ & $0.681$ & $4.74$ & $1.69$ & $3.12$ & $6.49$ & $4.11$ & $0.535$ & $0.953$ & $0.714$ & $3.60$ \\
        + \sixxtwo\ cons. & $3.80$ & $7.01$ & $0.724$ & $4.98$ & $1.81$ & $3.52$ & $6.92$ & $4.16$ & $0.552$ & $1.06$ & $0.770$ & $3.76$ \\
        \hline
        \rowcolor{gray!15} 
        \textbf{\CMBS} & $\times10^{-6}$& $\times10^{-5}$ & $\times10^{-3}$ & $\times10^{-3}$ & $\times10^{-11}$ & $\times10^{-3}$ & $\times10^{-3}$ & $\times10^{-3}$ & $\times10^{-2}$ & $\times10^{-3}$ & $\times10^{-1}$ & $\times10^{-1}$ \\
        CMB only & $2.94$ & $4.49$ & $2.10$ & $6.22$ & $2.29$ & $4.67$ & $7.10$ & $4.65$ & $1.04$ & $9.63$ & $7.24$ & $5.42$ \\
        + $C_\ell^{\kappa\kappa}$ & $2.78$ & $4.07$ & $1.56$ & $6.10$ &  $2.25$ & $4.00$ & $6.65$ & $4.37$ & $0.790$ & $8.94$ & $6.40$ & $5.03$ \\
        + \threextwo\ pess. & $2.38$ & $3.37$ & $0.653$ & $4.92$ & $1.76$ & $3.08$ & $6.01$ & $3.57$ & $0.491$ & $8.13$ & $5.72$ & $3.28$ \\
        + \threextwo\ cons. & $2.51$ & $3.39$ & $0.673$ & $5.18$ & $1.87$ & $3.36$ & $6.24$ & $3.59$ & $0.502$ & $8.62$ & $5.91$ & $3.38$ \\
        + \sixxtwo\ pess. & $2.31$ & $3.34$ & $0.577$ & $3.91$ & $1.33$ & $2.76$ & $5.65$ & $3.35$ & $0.464$ & $7.56$ & $5.29$ & $3.11$ \\
        + \sixxtwo\ cons. & $2.43$ & $3.36$ & $0.601$ & $4.13$ & $1.42$ & $3.09$ & $5.98$ & $3.39$ & $0.479$ & $8.20$ & $5.56$ & $3.21$ \\
    \hline
    \end{tabular}
    }
    \caption{Marginalized 1-$\sigma$ constraints for the indicated theoretical models and combination of probes. `pess' refers to the use of the \Euclid\ pessimistic scheme ($\ell_\textrm{GC}<750$, $\ell_\textrm{WL}<1500$) whereas `cons' is the even more conservative multipole range ($\ell_\textrm{GC}<500$, $\ell_\textrm{WL}<750$).\label{tab:errors}}
\end{table}
\clearpage

\section{Results}~\label{sec:results}

\begin{figure}[htbp]
\centering
\includegraphics[width=\textwidth]{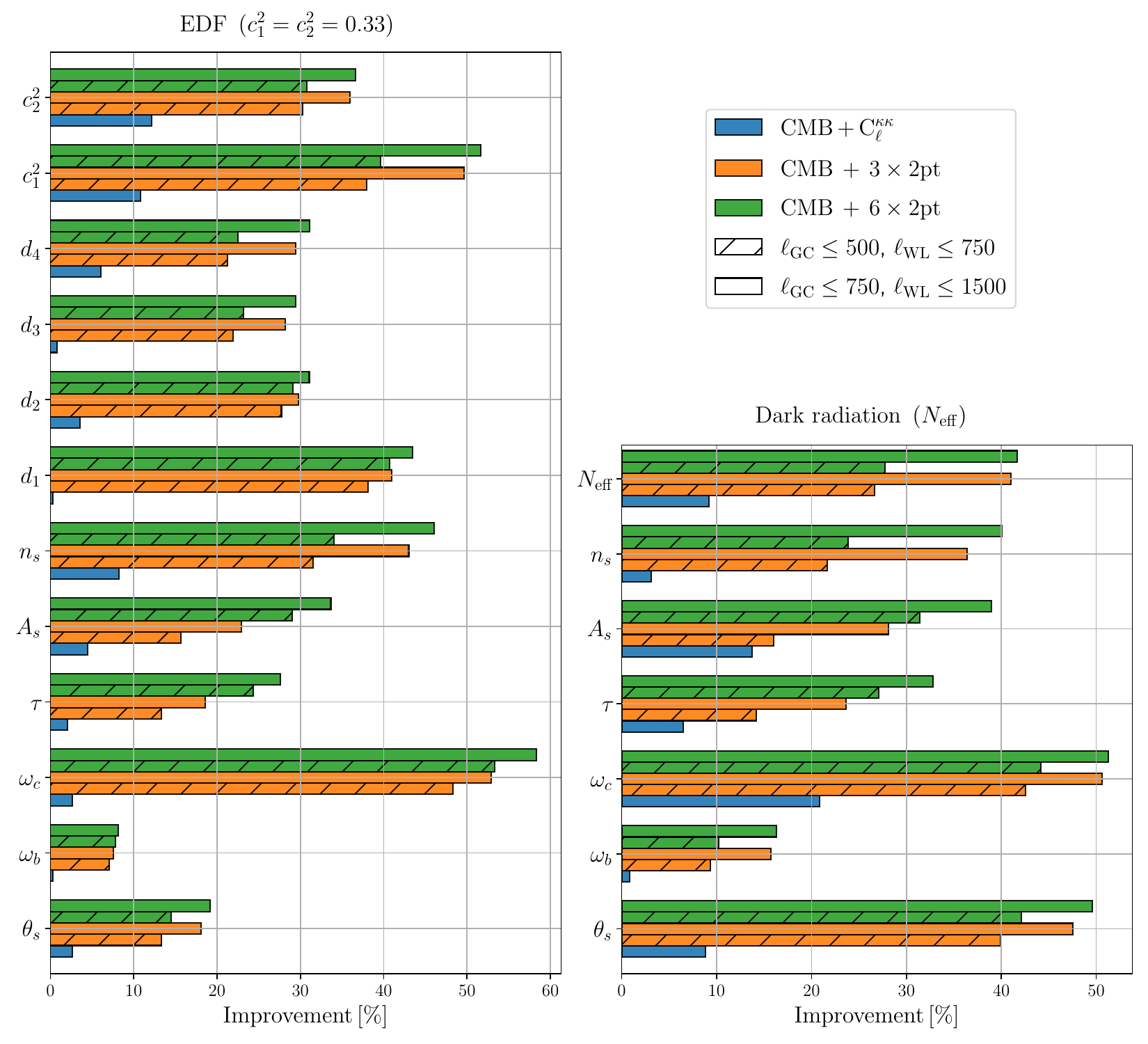}
\caption{Improvement in percentage on the 1-$\sigma$ constraint of the cosmological parameters in the case of EDF with $c_1^2=c_2^2=0.33$ (left) or in the \Neff\ case (right), when adding CMB lensing convergence power spectrum (blue), \threextwo\ power spectra (orange) or the full \sixxtwo\ power spectra (green) to the CMB primary power spectra observed by \SO. The plain bars correspond to the use of \Euclid\ pessimistic multipole ranges, while the dashed bars are obtained using an even more conservative multipole range.\label{fig:improvement_EDF_neff}}
\end{figure}

We compare the constraints that can be obtained on the three theoretical models we considered, depending on the probes used in the analysis. Table~\ref{tab:errors} shows the 1-$\sigma$ uncertainties that can be reached using different combinations of data, marginalized over all the nuisance parameters included in this forecast. In all cases, the addition of LSS data, and in particular of the \sixxtwo\ observables, leads to a significant tightening of the constraints. In the dark radiation case, constraints on \Neff\ shrink from $0.0646$ and $0.0397$ using the primary CMB observations from \SO\ and \CMBS, to $0.0377$ and $0.0250$ in the most optimistic scenario considered here. Figure~\ref{fig:improvement_EDF_neff} also shows in a more visual way the improvement that can be obtained for dark radiation models (either \Neff\ or EDF with $c_1^2=c_2^2=0.33$) by adding different combinations of LSS observations to \SO\ data, where the improvement is defined as
\begin{align}
    \textrm{Improvement} = 100\,\frac{\sigma_\textrm{CMB}(\alpha)-\sigma_\textrm{CMB+LSS}(\alpha)}{\sigma_\textrm{CMB}(\alpha)},
\end{align}
where $\alpha$ can be any cosmological parameter, and $\sigma_\textrm{CMB}(\alpha)$ and $\sigma_\textrm{CMB+LSS}(\alpha)$ denote the 1-$\sigma$ constraint that are obtained using only CMB data, or a combination of CMB and LSS data, respectively. The improvement we get when adding \Euclid\ data depends on the multipole ranges included in the analysis, showing that small scales are sensitive to \Neff. Modelling the small scales effects (nonlinearities, baryonic feedback) would therefore be crucial if one wanted to improve the constraints on \Neff\ by including even smaller scales. However, this forecast shows that using a very conservative multipole scheme (`cons.' in the table) can still bring most of the improvement, roughly at the level of $30\%$ compared to the use of the CMB primary power spectra alone. The improvement on most amplitude parameters in the EDF model, when the fiducial values of the sound speed parameters are $0.33$, similarly is approximately at the level of $30$--$40\%$ when adding the \sixxtwo\ observables. There is also a significant improvement on the sound speed parameters, $\sigma(c_1^2)$ in particular, being divided by about a factor of $2$ when adding LSS data.

\begin{figure}[htbp]
\centering
\includegraphics[width=\textwidth]{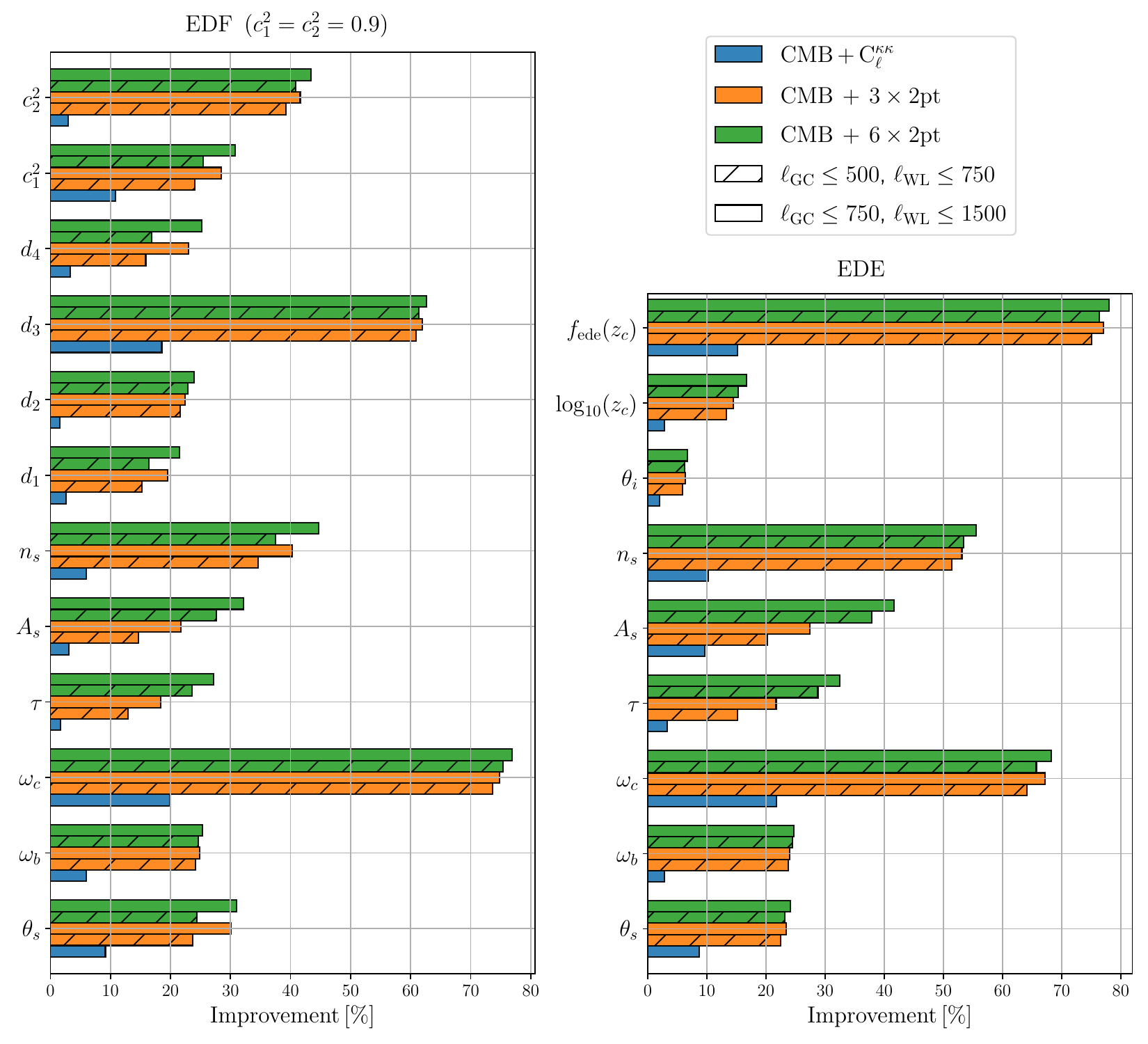}
\caption{Same as figure~\ref{fig:improvement_EDF_neff} in the case of EDF with $c_1^2=c_2^2=0.9$ (left) or EDE (right). \label{fig:improvement_EDF_EDE}}
\end{figure}

\begin{figure}[htbp]
\centering
\includegraphics[width=\textwidth]{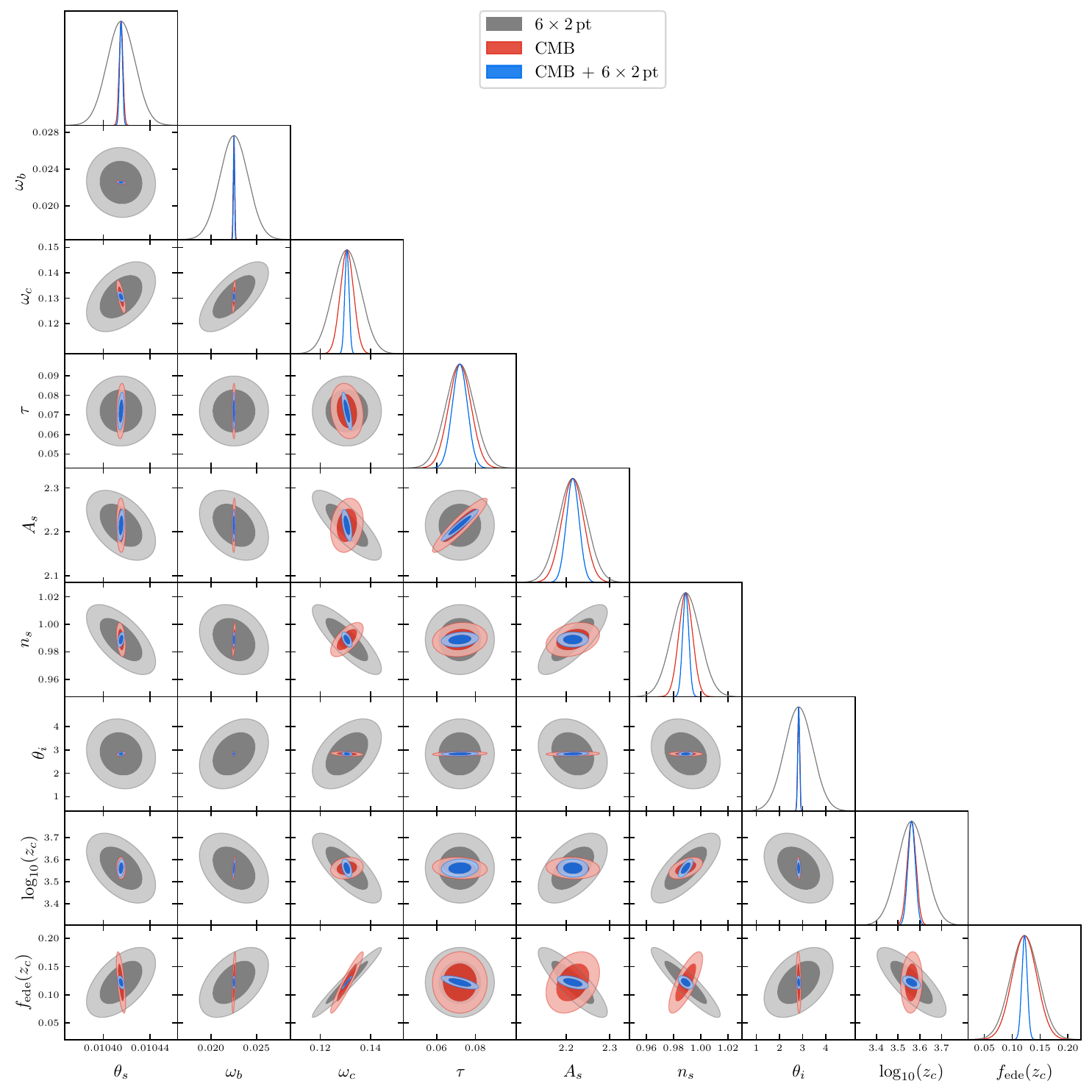}
\caption{Forecast constraints in the EDE case, using \SO\ and \Planck\ data only (red), the \sixxtwo\ observables from \Euclid\ and \SO\ with \Planck\ prior on $\tau$ (grey), or the combination of all those probes (blue). Combining CMB and LSS data will allow breaking degeneracies between parameters, especially when $A_s$, $n_s$ and $f_\textrm{ede}(z_c)$ are involved. \label{fig:triangle_plot_EDE}}
\end{figure}

In the case of EDE, or EDF with $c_1^2=c_2^2=0.9$, the improvement gained by adding LSS data is even more significant, as can be seen in table~\ref{tab:errors} and figure~\ref{fig:improvement_EDF_EDE}. The 1-$\sigma$ uncertainty of $f_\textrm{ede}$ is roughly divided by a factor of $3$ to $4$ in most cases. This result extends the findings from~\cite{Ivanov:2020ril}, who showed that  spectroscopic galaxy surveys would considerably improve constraints on EDE models. Interestingly, the improvement we find mainly comes from the largest scales, as the constraints do not change much when using the \Euclid\ pessimistic or the even more conservative multipole ranges. This shows that adding \Euclid\ data will lead to a massive and very robust improvement, as it will be possible to use large scales only, where the modelling is most trustworthy. In particular, large scales would not be affected by small scales systematics and matter power spectrum predictions in the linear or mildly non-linear regimes are very robust. Additionally, the use of a linear galaxy bias would be well justified and should not break down at those scales. Figure~\ref{fig:triangle_plot_EDE} shows the correlations between all the theoretical parameters when using the \sixxtwo\ observables, with \Planck\ prior on $\tau$ (grey), CMB primary power spectra (red) or the combination of the two (blue) in the case of \SO. It can be seen that combining those probes will be very efficient at breaking degeneracies, especially between $f_\textrm{ede}$ and $n_s$, leading to a massive reduction of the 1-$\sigma$ uncertainty in $f_\textrm{ede}$. The impact of EDE models on the CMB and matter power spectra, as well as the reasons behind the differing degeneracy directions, is well described in~\cite{Poulin23}. Adding EDE shifts the CMB acoustic peaks, an effect that is partially counteracted by an increase in $H_0$. However, EDE also causes gravitational potentials to decay more rapidly around recombination, enhancing the early ISW effect in the temperature power spectrum. To mitigate this, the dark matter density increases, preventing gravitational potentials from decaying too quickly. Once $H_0$ and $\omega_c$ are fixed, the remaining effect of EDE (relative to \lcdm) is a tilt in the power spectra, due to the stronger effect of EDE on the angular sound horizon than on the damping scale. A higher spectral index $n_s$ is then needed to balance this tilt.

In large-scale structure observables, EDE suppresses power because it induces a faster decay of the gravitational potentials which reduces perturbations growth around early matter domination, and because it does not cluster efficiently. This lack of power is counteracted by an increase in $\omega_c$, (which can be substantial, since it is not constrained by early ISW as it is the case in CMB observations). This increase of $\omega_c$ results in a tilt of the matter power spectrum, which can be compensated by a decrease of $n_s$.

Likewise, in the case where $c_1^2=c_2^2=0.9$, the parameter $d_3$ in the EDF model will be much better constrained when adding LSS data to CMB observations, as shown on figure~\ref{fig:improvement_EDF_EDE}, with an improvement of over $60\%$. The fact that the constraint on $d_3$ becomes significantly tighter with LSS data, similarly to the EDE case, suggests that the parameter $d_3$ leaves a similar imprint on the matter power spectrum as EDE. This finding is particularly interesting since~\cite{Kou:2024rvn} showed that this parameter was the most useful to reproduce the effect of EDE on the CMB power spectra. Therefore, the EDF framework not only is interesting to test early physics with CMB data, but also with upcoming galaxy surveys.
In the case of dark radiation models such as \Neff, we find that \Neff\ exhibits similar degeneracy directions with the \lcdm\ parameters as $f_\textrm{ede}$ does in the EDE case. This is specifically true for the degeneracies between \Neff\ and $A_s$, and between \Neff\ and $n_s$, stemming from the same physical mechanisms discussed above. Nonetheless, since future CMB data will already constrain \Neff\ very tightly (primarily through phase shifts and damping tail measurements), the additional constraining power from LSS we forecast will be more limited than in the EDE case.

We estimate the degradation due to the inclusion of the lensing covariance in the 1-$\sigma$ uncertainty of each parameter following

\begin{align}
    \textrm{Degradation} = 100\,\frac{\sigma_\textrm{tot}(\alpha)-\sigma_\textrm{Gauss}(\alpha)}{\sigma_\textrm{Gauss}(\alpha)},
\end{align}
where $\sigma_\textrm{tot}(\alpha)$ and $\sigma_\textrm{Gauss}(\alpha)$ respectively denote the 1-$\sigma$ uncertainties estimated in a parameter $\alpha$, using the total covariance (\textit{i.e.} including the lensing covariance) and the Gaussian covariance only. Table~\ref{tab:degradation} shows the degradations for all the parameters, for the different models and combination of experiments. In some cases, we find negative degradations, which means that including the lensing covariance can sometimes tighten the constraints instead of degrading them. This happens in particular because we are taking into account the cross-covariance between CMB and LSS observables, and that including the correlation between those observables can reduce the uncertainty in a parameter. This is in particular the case if the two observables exhibit a positive correlation and react to the parameter in opposite ways, or conversely, if they have a negative correlation but react to the parameter in the same way. 

In the models we studied, it appears that the degradation due to the lensing covariance is a small correction for most parameters, except for $\tau$ and $A_s$. Specifically, we find that including the lensing covariance can lead to a degradation of almost $12\%$ and $18\%$ for $\tau$ and $A_s$ in the EDE model, when considering a combination of \Euclid\ \threextwo\ observables with \CMBS. Interestingly, this degradation reduces considerably (by a factor of about $2$) when adding the CMB lensing convergence power spectrum. More generally, we find in all cases that including the CMB lensing power spectrum in the forecast reduces the degradation due to the lensing covariance. When looking at the early physics parameters, we find negligible impact for \Neff\ and EDE models in all the scenarios we considered. In the EDF model, we find a slight degradation for $c_1^2$ in the dark radiation-like case (up to $6.5\%$ for \CMBS) and to a lesser extent to $d_4$ in the quintessence-like case (up to $4.4\%$ for \CMBS). As could be expected, the impact of the lensing covariance is even more limited for \SO\ than it is for \CMBS, because of the higher noise level in \SO.

\begin{table}[ht]
    \centering
    \resizebox{0.85\textwidth}{!}{
    \begin{tabular}{lcccccccccccc}
    \hline
        \rowcolor{Emerald!30}
        \multicolumn{13}{c}{\textbf{Dark radiation (\Neff)}} \\
        \hline
        & $\theta_s$ & $\omega_b$ &$\omega_c$ & $\tau$ & $A_s$ & $n_s$ & \Neff & & & & & \\
        \hline
        \rowcolor{gray!15} 
        \textbf{\SO} & & & & & & & & & & & & \\
        CMB only & $1.3$ & $0.1$ & $2.6$ & $3.7$ & $5.9$ & $0.1$ & $0.5$ & & & & & \\
        + $C_\ell^{\kappa\kappa}$ & $-0.3$ & $0.6$ & $-0.3$ & $1.9$ & $2.2$ &$0.7$ & $0.3$ & & & & & \\
        + \threextwo\ pess. & $0.7$ & $0.6$ & $1.0$& $3.5$ & $3.6$ & $2.2$ & $1.6$ & & & & & \\
        + \threextwo\ cons. & $0.2$ & $0.4$ & $-0.1$ & $3.7$ & $3.6$& $1.2$&$0.5$ & & & & & \\
        + \sixxtwo\ pess. & $0.6$&$0.6$ &$1.2$ & $3.2$& $3.4$&$1.4$ &$1.5$ & & & & & \\
        + \sixxtwo\ cons. & $0.3$ & $0.6$& $0.6$& $3.0$&$3.1$ & $0.9$& $0.9$& & & & & \\
        \hline
        \rowcolor{gray!15} 
        \textbf{\CMBS} & & & & & & & & & & & & \\
        CMB only & $2.9$ & $0.1$ & $3.2$ & $5.7$ & $10.5$ & $0.1$ & $2.2$ & & & & & \\
        + $C_\ell^{\kappa\kappa}$ & $0.1$ & $0.6$ & $3.9$ & $3.8$ & $4.3$ & $0.8$ & $2.0$ & & & & & \\
        + \threextwo\ pess. & $0.2$ & $0.2$ & $-0.2$ & $8.1$ & $10.4$ & $2.3$ & $0.5$ & & & & & \\
        + \threextwo\ cons. & $-0.1$ & $0.0$ & $-0.9$ & $9.0$ & $11.5$ & $0.6$ & $-0.4$ & & & & & \\
        + \sixxtwo\ pess. & $0.2$ & $0.6$ & $1.8$ & $2.7$ & $3.4$ & $1.4$ & $2.0$ & & & & & \\
        + \sixxtwo\ cons. & $0.2$ & $0.8$ & $1.8$ & $2.2$ & $2.7$ & $1.3$ & $1.9$ & & & & & \\
        \hline
        \rowcolor{Emerald!30}
        \multicolumn{13}{c}{\textbf{Early Dark Energy}} \\
        \hline
        & $\theta_s$ & $\omega_b$ &$\omega_c$ & $\tau$ & $A_s$ & $n_s$ & $\theta_i$ & $\log_{10}{(z_c)}$ & $f_\textrm{ede}(z_c)$ & & & \\
        \hline
        \rowcolor{gray!15} 
        \textbf{\SO} & & & & & & & & & & & & \\
        CMB only & $0.7$& $0.1$& $1.1$& $2.8$&$5.3$ &$0.0$ & $0.0$& $0.1$&$0.3$ & & & \\
        + $C_\ell^{\kappa\kappa}$ & $0.1$ & $0.8$ & $0.3$ & $2.5$ & $2.5$ & $0.3$ & $0.4$ & $0.0$ & $0.5$ & & & \\
        + \threextwo\ pess. & $0.0$ & $0.1$ & $0.2$ & $4.5$ & $5.7$ & $0.7$ & $0.3$ & $0.2$ & $0.3$ & & & \\
        + \threextwo\ cons. & $0.0$ & $0.1$ & $-0.1$ & $5.6$& $7.0$ & $0.5$ & $0.3$ & $0.2$ & $-0.1$ & & & \\
        + \sixxtwo\ pess. & $0.0$ & $0.0$ & $0.4$ & $4.2$ & $6.1$ & $0.0$ & $0.2$ & $0.1$ & $0.4$ & & & \\
        + \sixxtwo\ cons. & $0.0$ & $0.0$ & $0.2$ & $4.7$ & $7.0$ & $-0.1$ & $0.2$ & $0.1$ & $0.2$ & & & \\
        \hline
        \rowcolor{gray!15} 
        \textbf{\CMBS} & & & & & & & & & & & & \\
        CMB only & $0.5$ & $0.0$ & $0.7$ & $4.7$ & $8.0$ & $0.1$ & $0.4$ & $0.3$ & $0.4$ & & & \\
        + $C_\ell^{\kappa\kappa}$ & $1.2$ & $0.5$ & $5.5$ & $5.0$ & $5.0$ & $-0.7$ & $1.7$ & $-0.1$ & $2.8$ & & & \\
        + \threextwo\ pess. & $0.2$ & $0.1$ & $0.3$ & $11.0$ & $16.3$ & $1.4$ & $0.9$ & $0.2$ & $0.3$ & & & \\
        + \threextwo\ cons. & $0.2$ & $0.1$ & $-0.2$ & $11.8$ & $17.8$ & $0.9$ & $1.0$ & $0.1$ & $-0.1$ & & & \\
        + \sixxtwo\ pess. & $0.0$ & $0.0$ & $0.9$ & $4.2$ & $7.1$ & $-1.2$ & $0.1$ & $-0.4$ & $1.0$ & & & \\
        + \sixxtwo\ cons. & $0.1$ & $0.0$ & $1.2$ & $4.6$ & $7.5$ & $-1.2$ & $0.2$ & $-0.5$ & $1.3$ & & & \\
        \hline
        \rowcolor{Emerald!30}
        \multicolumn{13}{c}{\textbf{Early Dark Fluid ($c_1^2=c_2^2=0.33$, $d_1=d_2=d_3=d_4=0.01$)}} \\
        \hline
        & $\theta_s$ & $\omega_b$ &$\omega_c$ & $\tau$ & $A_s$ & $n_s$ & $d_1$ & $d_2$ & $d_3$ & $d_4$ & $c_1^2$ & $c_2^2$ \\
        \hline
        \rowcolor{gray!15} 
        \textbf{\SO} & & & & & & & & & & & & \\
        CMB only & $0.4$ & $0.2$ & $0.9$ & $1.8$ & $3.5$ & $0.0$ & $0.1$ & $0.7$ & $0.0$ & $0.0$ & $0.0$ & $0.3$ \\
        + $C_\ell^{\kappa\kappa}$ & $1.3$ & $0.0$ & $0.2$ & $2.2$ & $2.7$ & $1.5$ & $0.1$ & $-0.2$ & $-0.1$ & $0.8$ & $2.0$ & $1.1$ \\
        + \threextwo\ pess. & $0.7$ & $0.0$ & $0.2$ & $2.7$ & $3.6$ & $2.0$ & $0.0$ & $0.0$ & $0.0$ & $0.6$ & $2.4$ & $1.0$ \\
        + \threextwo\ cons. & $0.9$ & $0.0$ & $0.5$ & $3.1$ & $3.8$ & $1.7$ & $0.2$ & $0.0$ & $0.0$ & $0.8$ & $2.3$ & $1.0$ \\
        + \sixxtwo\ pess. & $0.5$ & $0.0$ & $0.3$ & $2.0$ & $3.1$ & $1.0$ & $0.1$ & $-0.2$ & $0.0$ & $0.4$ & $2.0$ & $0.8$ \\
        + \sixxtwo\ cons. & $0.7$ & $0.0$ & $0.5$ & $2.1$ & $3.0$ & $0.9$ & $0.2$ & $-0.3$ & $-0.1$ & $0.5$ & $2.0$ & $0.9$ \\
        \hline
        \rowcolor{gray!15} 
        \textbf{\CMBS} & & & & & & & & & & & & \\
        CMB only & $3.8$ & $0.0$ & $3.8$ & $5.1$ & $7.0$ & $1.3$ & $2.8$ & $0.0$ & $0.0$ & $1.3$ & $3.3$ & $0.7$ \\
        + $C_\ell^{\kappa\kappa}$ & $3.7$ & $0.0$ & $-0.7$ & $5.8$ &  $6.2$ & $2.1$ & $-0.1$ & $-1.7$ & $-0.4$ & $0.8$ & $6.5$ & $3.9$ \\
        + \threextwo\ pess. & $3.5$ & $0.0$ & $1.2$ & $7.8$ & $10.3$ & $3.7$ & $0.6$ & $0.0$ & $0.1$ & $1.9$ & $6.3$ & $2.2$ \\
        + \threextwo\ cons. & $3.5$ & $0.0$ & $1.6$ & $8.5$ & $10.7$ & $2.9$ & $1.1$ & $0.0$ & $0.2$ & $2.1$ & $5.7$ & $2.2$ \\
        + \sixxtwo\ pess. & $2.8$ & $0.0$ & $-0.5$ & $1.1$ & $2.2$ & $1.0$ & $-0.1$ & $-1.8$ & $-0.2$ & $-0.3$ & $2.5$ & $0.7$ \\
        + \sixxtwo\ cons. & $3.4$ & $0.0$ & $-0.4$ & $1.2$ & $2.0$ & $1.4$ & $0.0$ & $-2.2$ & $-0.8$ & $0.1$ & $3.6$ & $1.4$ \\
    \hline
    \rowcolor{Emerald!30}
        \multicolumn{13}{c}{\textbf{Early Dark Fluid ($c_1^2=c_2^2=0.9$, $d_1=d_2=d_3=d_4=0.01$)}} \\
        \hline
        & $\theta_s$ & $\omega_b$ &$\omega_c$ & $\tau$ & $A_s$ & $n_s$ & $d_1$ & $d_2$ & $d_3$ & $d_4$ & $c_1^2$ & $c_2^2$ \\
        \hline
        \rowcolor{gray!15} 
        \textbf{\SO} & & & & & & & & & & & & \\
        CMB only & $0.6$ & $0.0$ & $0.4$ & $1.1$ & $2.3$ & $1.2$ & $1.5$ & $1.7$ & $0.3$ & $0.4$ & $1.8$ & $1.4$ \\
        + $C_\ell^{\kappa\kappa}$ & $0.3$ & $1.4$ & $1.7$ & $1.2$ & $1.7$ & $2.2$ & $0.3$ & $1.6$ & $1.2$ & $1.6$ & $-0.1$ & $1.3$ \\
        + \threextwo\ pess. & $0.3$ & $0.0$ & $0.0$ & $3.0$ & $4.4$ & $0.7$ & $0.1$ & $0.0$ & $-0.2$ & $1.1$ & $-0.2$ & $-0.1$ \\
        + \threextwo\ cons. & $0.5$ & $0.0$ & $0.1$ & $3.4$ & $5.2$ & $0.4$ & $0.0$ & $0.0$ & $-0.2$ & $0.9$ & $-0.2$ & $-0.1$ \\
        + \sixxtwo\ pess. & $0.0$ & $0.0$ & $0.3$ & $2.4$ & $4.2$ & $0.1$ & $0.1$ & $-0.2$ & $-0.1$ & $-0.9$ & $-0.1$ & $0.0$ \\
        + \sixxtwo\ cons. & $0.6$ & $0.0$ & $0.4$ & $2.5$ & $4.7$ & $-0.1$ & $-0.1$ & $-0.3$ & $-0.2$ & $0.6$ & $-0.1$ & $0.1$ \\
        \hline
        \rowcolor{gray!15} 
        \textbf{\CMBS} & & & & & & & & & & & & \\
        CMB only & $0.4$ & $0.4$ & $0.1$ & $2.8$ & $3.8$ & $4.0$ & $2.9$ & $3.8$ & $0.0$ & $4.4$ & $3.3$ & $2.7$ \\
        + $C_\ell^{\kappa\kappa}$ & $0.4$ & $2.1$ & $6.4$ & $3.2$ &  $3.6$ & $1.3$ & $-0.8$ & $0.4$ & $2.4$ & $1.8$ & $-0.4$ & $1.0$ \\
        + \threextwo\ pess. & $0.4$ & $0.0$ & $0.2$ & $8.2$ & $12.2$ & $2.3$ & $1.1$ & $0.4$ & $0.0$ & $3.6$ & $0.4$ & $-0.1$ \\
        + \threextwo\ cons. & $0.5$ & $0.0$ & $0.2$ & $8.7$ & $13.0$ & $1.5$ & $0.7$ & $0.4$ & $-0.1$ & $2.9$ & $0.1$ & $-0.1$ \\
        + \sixxtwo\ pess. & $1.1$ & $0.0$ & $0.1$ & $1.6$ & $4.0$ & $-1.1$ & $-1.0$ & $-2.5$ & $-1.7$ & $-0.4$ & $-0.3$ & $0.1$ \\
        + \sixxtwo\ cons. & $1.6$ & $0.0$ & $0.1$ & $1.6$ & $4.0$ & $-1.0$ & $-0.9$ & $-2.5$ & $-1.7$ & $0.1$ & $-0.2$ & $0.1$ \\
    \hline
    \end{tabular}
    }
    \caption{Degradation (in percentage) due to the inclusion of the lensing covariance.}\label{tab:degradation}
\end{table}

\section{Conclusion}~\label{sec:conclusion}

In this work, we study how LSS data from photometric stage IV surveys such as \Euclid\ will tighten constraints on early physics models, when combined with CMB observations from future experiments like \SO\ or \CMBS. We consider several early physics extensions to \lcdm, namely the addition of extra relativistic species through \Neff, the axion-like EDE, or the flexible EDF model with two configurations mimicking a dark radiation or a quintessence field. We model the covariance that arises from the gravitational lensing induced by the large-scale structure of the Universe. This has an impact on the CMB covariance by introducing correlations between different modes. It also introduces a cross-covariance between CMB and LSS tracers. We find that for the combination of \Euclid\ and either \SO\ or \CMBS, the impact of this lensing induced covariance on the cosmological constraints is rather small for the models we tested, apart for $A_s$ and $\tau$, where taking into account this covariance degrades the constraints by $10$--$20\%$.

Overall, we find that adding LSS data would lead to significant improvements, of the order of $30$--$40\%$ for \Neff, to almost $80\%$ in the EDE case. Interestingly, much of the improvement comes from the observation of LSS tracers on large scales, whose modelling is more robust and is less affected by systematic effects. This is especially true in the EDE case, where almost all the constraining power comes from the largest scales. This is because early physics modifies the growth of perturbations around radiation-matter equality, leaving an imprint on the matter power spectrum at large scales, around \keq. Those signatures of the early physics allow breaking degeneracies between \lcdm\ and early physics parameters, as was shown on figure~\ref{fig:triangle_plot_EDE}. These results highlight the opportunity offered by upcoming CMB and LSS surveys to test deviations from \lcdm\ and probe the physics of the early Universe with greater precision.

\appendix
\section{Lensing-induced covariance of \SO}~\label{appendix:SO_cov}

\begin{figure}[htbp]
\centering
\includegraphics[width=\textwidth]{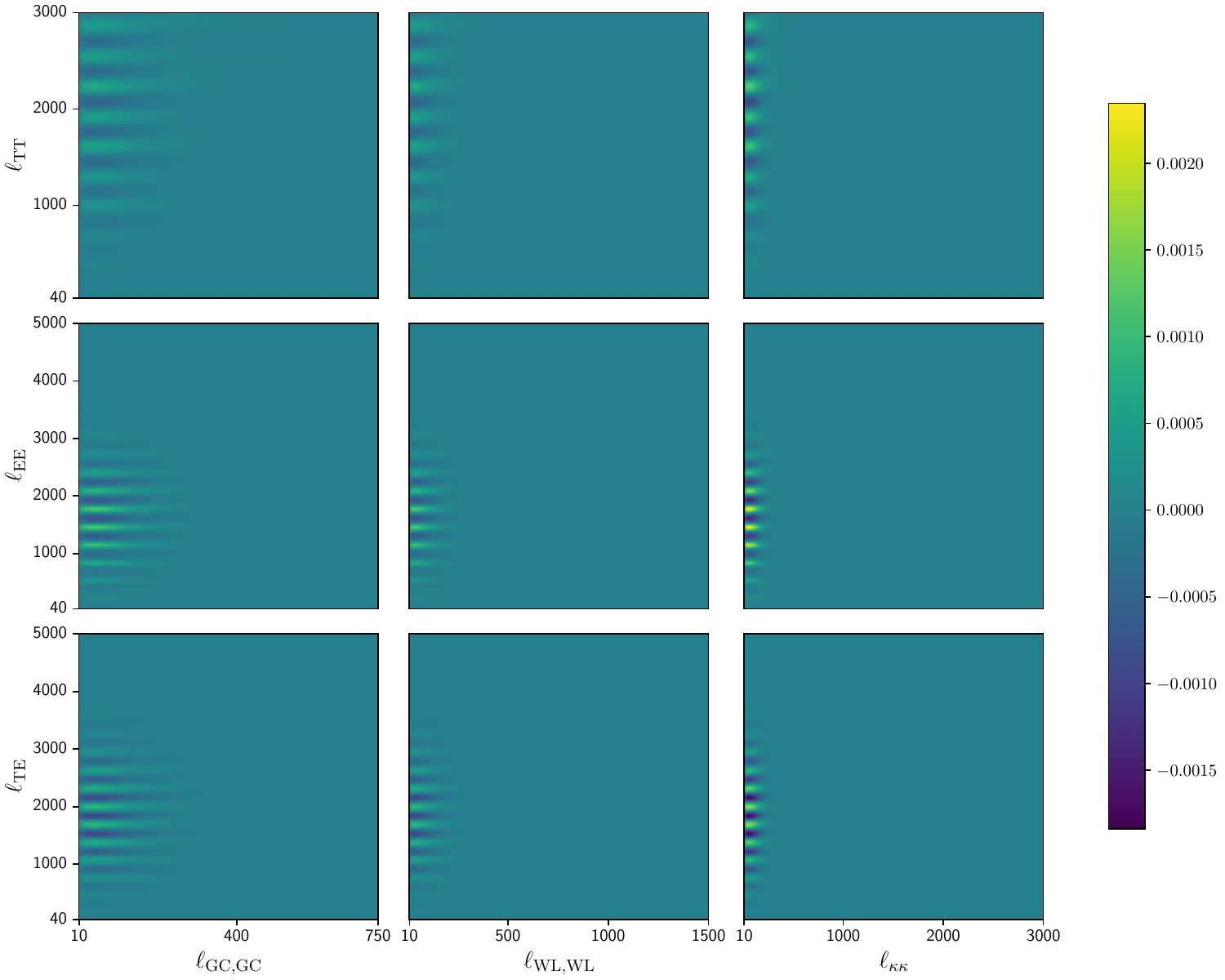}
\caption{Same as figure~\ref{fig:cross_covariance} in the case of \SO.\label{fig:cross_covariance_SO}}
\end{figure}

\begin{figure}[htbp]
\centering
\includegraphics[width=\textwidth]{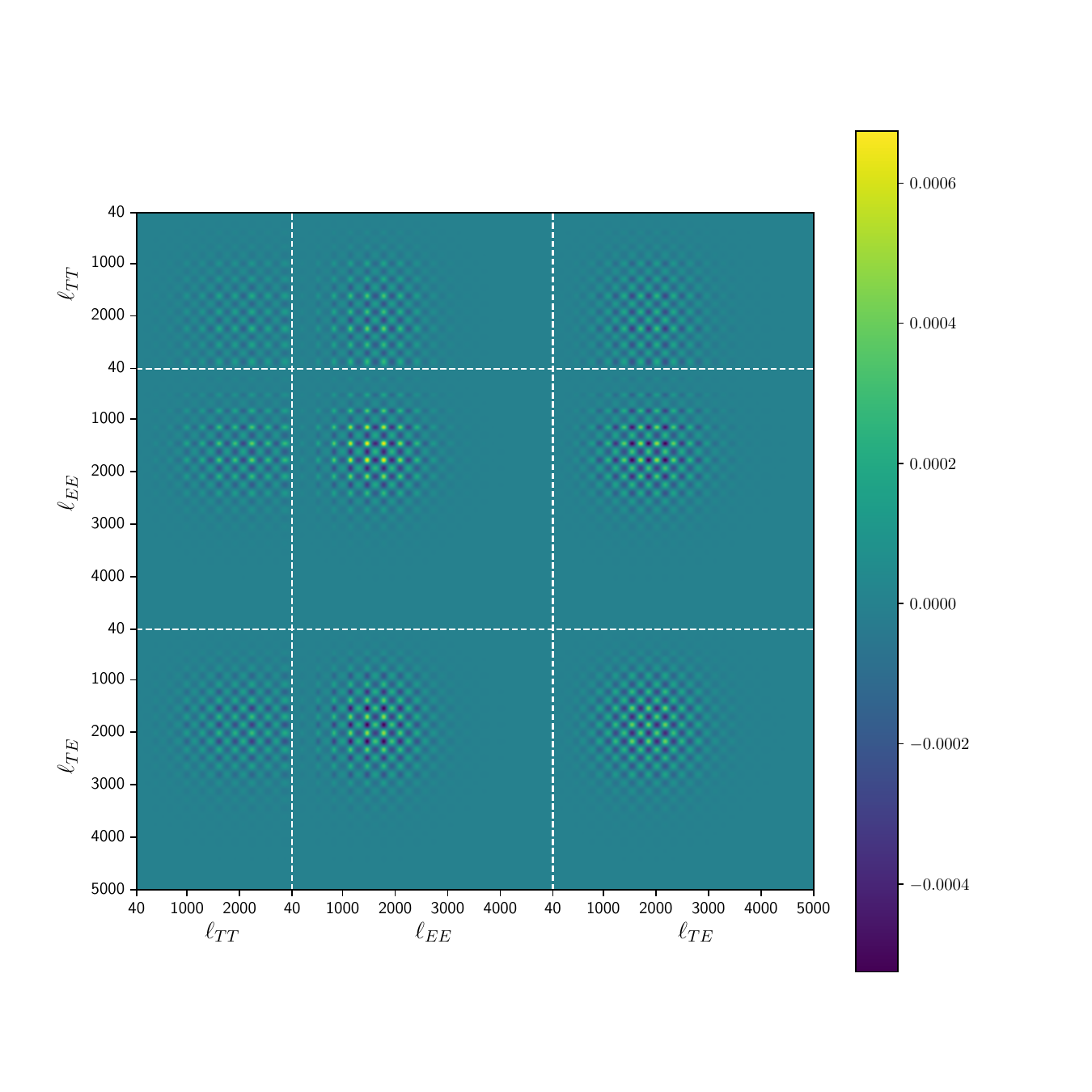}
\caption{Same as figure~\ref{fig:cov_CMB_lensing} in the case of \SO.\label{fig:cov_CMB_lensing_SO}}
\end{figure}

\acknowledgments

We are supported by UK STFC grant ST/X001040/1. We thank members of the \SO\ and \Euclid\ collaborations for discussion.

\bibliographystyle{JHEP}
\bibliography{biblio}

\end{document}